\documentclass[a4paper,12pt,UTF8]{article}

\usepackage{authblk}
\usepackage{amsmath,amsthm,amssymb}
\usepackage{algorithm}  
\usepackage{algpseudocode} 
\usepackage{dsfont}
\usepackage{appendix}
\usepackage{graphicx}%插入图片
\usepackage{float}%固定图片的位置
\graphicspath{{figures/}}%文章所用图片在当前目录下的 figures目录
\usepackage{color}
\usepackage{dsfont}
\usepackage{epstopdf}
\usepackage[hang]{subfigure}%多个图片共有
\usepackage{setspace}

\usepackage{natbib} %控制参考文献格式的代码
\usepackage{mathrsfs}% 不带花体小写字母
\usepackage{cases}
\usepackage{bm}
\usepackage{systeme}
\usepackage{ulem}
\usepackage{color}
\usepackage{textcomp}
\usepackage{amsfonts,amssymb}
\usepackage{multirow}
\usepackage{multicol}
\usepackage{enumerate}

\usepackage{xr}
\usepackage{booktabs}
\usepackage{threeparttable}
\usepackage{xcolor}
\usepackage{caption}
\usepackage{multirow}

\newtheorem{theo}{Theorem}[section]

\newtheorem{defi}{Definition}[section]
\newtheorem{prop}{Proposition}[section]

\RequirePackage[colorlinks,citecolor=blue,urlcolor=blue]{hyperref}
\renewcommand\emph[1]{{\color{red}\itshape #1}}

\newcommand\Rm[1]{\uppercase\expandafter{\romannumeral#1}}
\def\lam{\lambda}

\def\bbeta{{\bm\beta}}
\def\diag{\bm {\mbox{diag}}}

\newcommand\norm[1]{\Vert #1 \Vert}

\newcommand\inprod[1]{\langle #1 \rangle}

\def\defby{\stackrel{\mbox{\textrm{\tiny def}}}{=}}
\def\mR{\mathbb{R}}

\newcommand\mI{\mathbb{I}}

\def\bS{\mathbf{S}}

\def\tr{\mbox{tr}}

\def\a{{\bf a}}
\def\b{{\bf b}}
\def\A{{\bf A}}

\def\P{{\bf P}}

\def\I{{\bf I}}

\def\W{{\bf W}}
\def\I{{\bf I}}
\def\X{{\bf X}}
\def\x{{\bf x}}

\def\y{{\bf y}}

\def\defby{\stackrel{\mbox{\textrm{\tiny def}}}{=}}
\newcommand{\trans}{^{\mbox{\tiny{T}}}}

\newcommand{\bSig}{\mbox{\boldmath $\Sigma$}}

\newcommand{\bdelta}{\mbox{\boldmath $\delta$}}

\newcommand{\bPsi}{\mbox{\boldmath $\Psi$}}

\newcommand{\hbeta}{\mbox{$\widehat{\bbeta}$}}

\newcommand{\bxi}{\mbox{\boldmath $\xi$}}
\newcommand{\bpsi}{\mbox{\boldmath $\psi$}}
\newcommand{\bfeta}{\mbox{\boldmath $\eta$}}

\numberwithin{equation}{section}

\title{Distributed Reconstruction from Compressive Measurements: Nonconvexity and Heterogeneity}
\author{Erbo Li, Qi Qin, Yifan Sun, Liping Zhu}

\begin{document}

\def\spacingset#1{\renewcommand{\baselinestretch}%
{#1}\small\normalsize} \spacingset{1}

\maketitle

\bigskip
\begin{abstract}
The compressive sensing (CS) and 1-bit CS have gained prominence for their efficiency in signal acquisition and computational resource conservation. By utilizing only sign information, 1-bit CS achieves superior resource savings compared to conventional CS. With the emergence of massive data, the distributed signal aggregation under CS and 1-bit CS measurements introduces many challenges, including nonconvexity and heterogeneity. The nonconvexity originates from signal magnitude alignment operations designed to resolve the unidentifiability under finite-precision measurements. The heterogeneity arises from the discrepancies of the signal, noise intensities, and sign-flip probabilities on each node. 
To address these challenges, we propose a novel optimization framework with a squared cosine similarity penalty for robustness in heterogeneity. For intractable nonconvexity, our innovative solution leverages an invex relaxation method, which guarantees both uniqueness and algorithmic convergence to global optimality.
For signal and noisy measurement heterogeneity, the proposed estimator can be formulated as an adaptively de-biased least squares solution, where the correction leverages information from similar nodes, with  strength adaptively adjusted according to similarity and signal-to-noise ratio (SNR). As long as a certain level of signal similarity, under 1-bit CS measurements, our method achieves a high probability convergence rate improved from $O\{(p\log{p}/n_j)^{1/2}\}$ to $O\{(p\log{p}/N)^{1/2}+p^{1/2}/n_j\}$, where $p$ is the dimension of signals, $n_j$ and $N$ are local and pooling sample sizes respectively.
It can achieve the minimax optimal rate, with sufficient nodes of similarity. Numerical simulations validate the effectiveness and efficiency of our methods, demonstrating that our communication-efficient distributed algorithm enhances performance in reconstructing heterogeneous signals from 1-bit measurements. The proposed methods, algorithms, and theoretical results are also applicable to CS measurements. 
\end{abstract}

\spacingset{1.8}

\section{Introduction}\label{sec:intro}

Both compressive sensing (CS) \citep{donoho2006compressed} and 1-bit CS \citep{boufounos20081,knudson2016one} are pivotal methodologies that transcend the constraints of the Nyquist Sampling Theorem and enable robust recovery of signals from noisy and underdetermined measurements. The mathematical foundation of CS \citep{candes2006robust,candes2006stable} has inspired a series of statistical methods, including Lasso regression \citep{tibshirani1996regression} and the Uncertainty Autoencoder \citep{grover2019uncertainty}. To achieve remarkable efficiency in data storage and transmission, compared to CS, 1-bit CS \citep{boufounos20081,knudson2016one} measurements leverage sign measurements to quantize infinite-precision signals. However, extreme quantization introduces nonlinearity in 1-bit CS measurements, posing significant algorithmic and theoretical challenges. The signal becomes non-identifiable, though its direction can be reconstructed with high fidelity \citep{huang2018robust,knudson2016one}. 

The emergence of massive-scale data necessitates signal aggregation and reconstruction from both CS and 1-bit CS measurements in distributed settings with communication efficiency, which introduces challenges such as nonconvexity and heterogeneity. The nonconvexity comes from the magnitude alignment operations designed to resolve the unidentifiability inherent in 1-bit CS measurements. For instance, \cite{dai2016noisy} formulated the reconstruction problem as a nonconvex optimization problem with a unit sphere constraint. The challenging heterogeneity issue in distributed systems arises mainly from heterogeneous signals and noisy measurements. Let us consider the examples of wireless sensor networks, such as spatial and velocity radar arrays in autonomous vehicles \citep{liu2023bevfusion} or electrode sensor arrays in EEG systems \citep{pfurtscheller1997motor,haboba2011architecture}. These sensors receive spatial information, velocity data, or patterns with diverse characteristics. Meanwhile, wireless channels of nodes may suffer from different degrees of corruption, where various node qualities result in heterogeneous noisy measurements with different noise intensities and sign-flip probabilities \citep{chen2023distributed}. Despite the different orientations with node-specific perturbations or scaling, signals across nodes may share underlying patterns and latent similarity information, reflected as directional similarities. Ignoring potential directional similarities may discard valuable inter-node information, leading to suboptimal local estimates. Thus, striking a balance between heterogeneity and shared structure is critical for accurate distributed reconstruction.

To address the nonconvexity and heterogeneity challenges, we propose a collaborative framework that leverages directional alignment among signals in a distributed setting, where nodes exhibit heterogeneous signals and noisy measurements with various noise intensities and sign-flip probabilities. Specifically, we combine $\ell_2$ loss with a squared cosine similarity penalty to encourage consistency between signal directions while preserving node-specific adaptations. \cite{huang2018robust} have demonstrated that ordinary least squares (LS) can effectively address the reconstruction challenge of 1-bit CS. However, \citep{chen2023distributed} only extended their method to distinct node quality but homogeneous signal systems with sign-flips and node-variant noises. 

The squared cosine similarity penalty induces a nonconvex optimization landscape, though it is identifiable for 1-bit compressed signals. We resolve this nonconvexity through a novel technique \textit{invex relaxation}, which guarantees the uniqueness of the global minimum and enables efficient distributed optimization through gradient-based updates.
Unlike traditional convex relaxations \citep{gu2024robust}, our invex relaxation provides the global optimality of the unique Karush-Kuhn-Tucker (KKT) point while preserving the complete gradient information of the nonconvex penalty term within the algorithm. 
Leveraging the complete gradient information, we provide an analytical insight that elucidates the proposed method, improving estimation performance by jointly identifying similarity and noise intensity, thereby adaptively incorporating information from similar nodes.

Despite the heterogeneity of signals and noisy measurements, the theoretical results demonstrate that our method can adaptively identify the similarities among signals and the signal-to-noise ratio (SNR) of each node.
Specifically, our proposed estimator can be formulated as LS with corrections from other nodes. Here, similar nodes contribute higher-intensity debiasing terms, whereas heterogeneous nodes produce attenuated correction terms.
Under significant noise and sign-flip perturbations, the estimator intensifies correction components for local information refinement through SNR-driven adaptation. 

Provided that a sufficient degree of inter-node signal similarity exists, our method derives a tighter error bound than local least squares and improves the high probability global convergence rate, from $O\{(p\log{p}/n_j)^{1/2}\}$ to $O\{(p\log{p}/N)^{1/2}+p^{1/2}/n_j\}$, where $p$ is the dimension of signals, $n_j$ and $N$ are local and total sample sizes, respectively. With sufficient numbers of similar nodes, the estimate on each node achieves the minimax optimal rate for $\ell_2$-loss in $\mR^p$ globally with high probability. The theoretical framework presented in this work establishes that the proposed communication-efficient algorithm converges to the global optimum with minimax optimal statistical convergence rates.

To the best of our knowledge, this work represents the first attempt to address the reconstruction of 1-bit CS signals in the distributed setting with heterogeneous signals and variable nodes. Although our methodology development and theoretical analysis focus exclusively on the more challenging 1-bit CS measurement, the methodological framework, computational algorithms, and theoretical results developed in this study remain equally applicable to CS measurements.

Our main contributions are summarized as follows:
\begin{enumerate}

\item We propose a method incorporating $\ell_2$ loss with a squared cosine similarity penalty to aggregate heterogeneous signals in distributed scenarios. To address the nonconvexity introduced by this formulation, we develop an innovative invex relaxation framework, which provides the global optimality of the unique KKT. This relaxation framework reserves complete gradient information of the original problem and enables communication-efficient distributed optimization through gradient-based updates.
\item To address signal heterogeneity and noisy measurements, our theoretical results demonstrate that our method can be regarded as a corrected LS estimator. This estimator automatically identifies both the degree of inter-node similarity and the SNR, thereby adaptively incorporating relevant information to enhance local estimation accuracy. As long as a certain level of similarity exists between nodes, our method adaptively leverages information from similar nodes, significantly improving the algorithm’s convergence rate to minimax optimal rate for $\ell_2$-loss in $\mR^p$ globally.
\item 
For the proposed method and algorithm, we establish theoretical guarantees addressing nonconvexity and heterogeneity, with the detailed reasoning outlined as follows.
In Theorem \ref{Theorem:0}, we obtain the uniqueness of the global optimum and the correspondence between KKT points and global optima within the invex relaxation framework. This theoretical foundation enables us to design a communication-efficient distributed algorithm. 
Theorem \ref{Theorem:1} provides a tighter error bound and shows that, as long as the lower bound for squared cosine similarity satisfies $|\cos\theta| > O\{(p\log{p}/n_j)^{1/2}\}$,  our proposed method adaptively leverages information from similar nodes, significantly improving the algorithm convergence rate, from $ O\{(p\log{p}/n_j)^{1/2}\}$ to $ O\{(p\log{p}/N)^{1/2}+p^{1/2}/n_j\}$ with a high probability, for node $j=1,..,m$.
Additionally, Theorem \ref{Theorem:2} proves that, by transmitting gradients to solve the corresponding projected gradient descent problem, our algorithm converges to the global optimum, independent of initialization.
\end{enumerate}

The rest of this article is organized as follows. The methodology development is presented in Section  \ref{sec:method} and the theoretical guarantees are provided in Section \ref{sec:theorem}. In Section \ref{sec:simulation}, we present numerical simulations that validate the effectiveness and efficiency of our methods. Then, we show an application of our method to reconstruct EEG signals under 1-bit CS measurement using the SEED dataset in Section \ref{sec:application}. Finally, in Section \ref{sec:conclusion}, we provide a concise summary and discussion of our work. 
All appendices are relegated to the supplementary materials.

\section{Methodology development}\label{sec:method}
\subsection{Problem setup}
Consider a standard distributed system with $m$ nodes, which could represent sensors in a wireless sensor network, along with a central server. In this architecture, the server maintains communication capability with each node under some predefined protocol. In the 1-bit CS setup, at the $j$-th node, the measurement model is given by: 
\begin{equation}
\label{eq:model}
\y_j=\bxi_j\odot\text{sign}(\X_j\trans\bbeta_j^*+\bm\varepsilon_j), j=1,\ldots,m,
\end{equation}
where $\y_j=(y_{1,j},\ldots,y_{n_j,j})\in\mR^{n_j}$ is the 1-bit measurement, $\bbeta_j^*\in\mR^p$ is the unknown parameter of interest, i.e., the signal to recover, which may vary across nodes, $\X_j=(\x_{1,j},\ldots,\x_{n_j,j})\in\mR^{p\times n_j}$ is the measurement matrix,  $\bxi_j\in \mR^{n_j}$ is a random vector with independent and identically distributed (i.i.d.) entries $\xi_{i,j}$ modeling the sign flip of $\y_j$ with $\text{Pr}(\xi_{i,j}=1)=1-\text{Pr}(\xi_{i,j}=-1)=q_j\neq \frac{1}{2}$, and $\bm\varepsilon_j$ is an independent random error with mean 0 and variance $\sigma_j^2$. Throughout this paper, $\text{sign}(\cdot)$ represents the sign function with $\text{sign}(z)=1$ if $z\geq 0$ and $-1$ otherwise, and $\odot$ is the point-wise Hadamard product. We focus on the overdetermined setting where $n_j>p$. Due to the unknown variance parameter $\sigma_j$ in the model (\ref{eq:model}) is unidentifiable \citep{knudson2016one}. Consequently, our goal reduces to estimating the direction rather than its magnitude, as the parameter vector can only be recovered up to an unknown scaling constant. In addition, we use $\tr(\A)$ to represent the trace of the matrix $\A$. The notation $O(\cdot)$ represents the order up to a constant factor, where $a_n = O(b_n)$ denotes $|a_n/b_n|\le c\le\infty$ for a certain constant $c$ when $n$ is large enough.

\subsection{Nonconvex aggregation}
\cite{huang2018robust} have shown that the direction of $\bbeta_j$ can be estimated from 1-bit measurements by minimizing the quadratic loss function ${L}_j(\bbeta_j):=\|\y_j-\X_j\trans\bbeta_j\|^2/n_j$ at each local node. 

As discussed in the Section \ref{sec:intro}, in many real-world distributed 1-bit CS scenarios, signals captured by different nodes exhibit a certain degree of angular similarity despite their distinct characteristics. To make full use of this inherent similarity, we estimate all $\bbeta_j$s simultaneously using measurements scattered across different nodes. Specifically, we consider the following optimization problem: 
\begin{equation}
\min_{\bbeta}G_\lam(\bbeta) = \sum_{j=1}^m G_{\lam,j}(\bbeta) , \quad {\rm where}\quad G_{\lam,j}(\bbeta)\defby \left[{L}_j(\bbeta_j)-\frac{\lambda}{2m}\sum_{k=1}^m \cos^2\inprod{\bbeta_j,\bbeta_k}\right], \label{Formulation:OriginalProblem}
\end{equation}
where $\bbeta=(\bbeta_1\trans,\ldots,\bbeta_m\trans)\trans$, $\lambda$ is a regularization parameter to balance the individual utilities and the similarity among signals $\bbeta_j$s, $\inprod{\a,\b}$ represents the angle between  the vectors $\a$ and $\b$, and $\cos\inprod{\bbeta_j,\bbeta_k}=\bbeta_j\trans\bbeta_k/\|\bbeta_j\|\|\bbeta_k\|$
is the cosine similarity between the $i$-th and the $j$-th node's signals. We use the global minimum of the problem \eqref{Formulation:OriginalProblem} to estimate $m$ signals, $\bbeta_j$ $(j=1,\ldots,m)$.

The objective function comprises two key components. The first term models the empirical loss at each node, while the second term measures pairwise signal similarity across nodes. We opt for cosine similarity over $l_2$ distance, i.e., $\|\bbeta_j-\bbeta_k\|_2$, for several reasons. Firstly, cosine similarity is insensitive to signal magnitude, focusing on orientation. This characteristic allows it to capture true similarities between signals even when their magnitudes differ significantly. In contrast, $l_2$ distance emphasizes magnitude differences, potentially overshadowing directional similarities.	
Secondly, by measuring angular differences, cosine similarity better identifies structural and pattern similarities among signals rather than merely their absolute positions. This makes it particularly suitable for scenarios where understanding the relationships and patterns among signals is crucial. Last but not least, cosine similarity naturally normalizes values to a fixed range, easing regularization parameter tuning across different scenarios. 

The square of cosine similarity serves two critical purposes. First, in 1-bit CS, noise or other factors can cause sign flips, making similar signals appear dissimilar. Squaring the cosine similarity accounts for these false reversals, ensuring that the objective function rewards true directional similarities even when the measurement signs are flipped. Second, unlike squared $l_2$ distance, which cannot handle falsely reversed directions, squaring cosine similarity penalizes dissimilarity more heavily, promoting a more accurate representation of underlying signal directions.

\subsection{A novel invex relaxation}
The introduction of a cosine similarity term renders the objective function nonconvex, significantly increasing the complexity of finding an optimal solution. This nonconvexity causes algorithms to be highly dependent on initial values; different initializations can lead to markedly distinct signal estimation outcomes when using standard gradient descent to minimize the objective function (see Table \ref{tab:initialization} in Appendix \ref{app:suppResult}). Theoretically, the sensitivity to initial values complicates the establishment of statistical properties for the final iteration results. Such analyses usually require assumptions about the distance between the initial guess and the true value, but these assumptions are often unverifiable in practice.

\begin{figure}[!htbp]
    % \flushleft %左对齐
    \centering
    \subfigure[The nonconvex objective function]{\label{Figure:NonconvexIllustration}
        \includegraphics[width =0.25\textwidth]{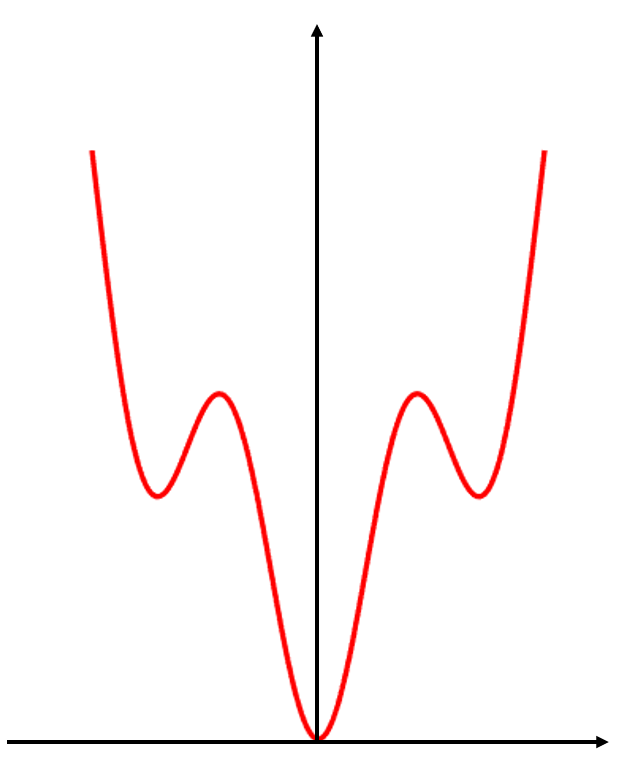}}
        \quad\quad\quad\quad
        \subfigure[The invex objective function]{\label{Figure:InvexIllustration}
        \includegraphics[width =0.25\textwidth]{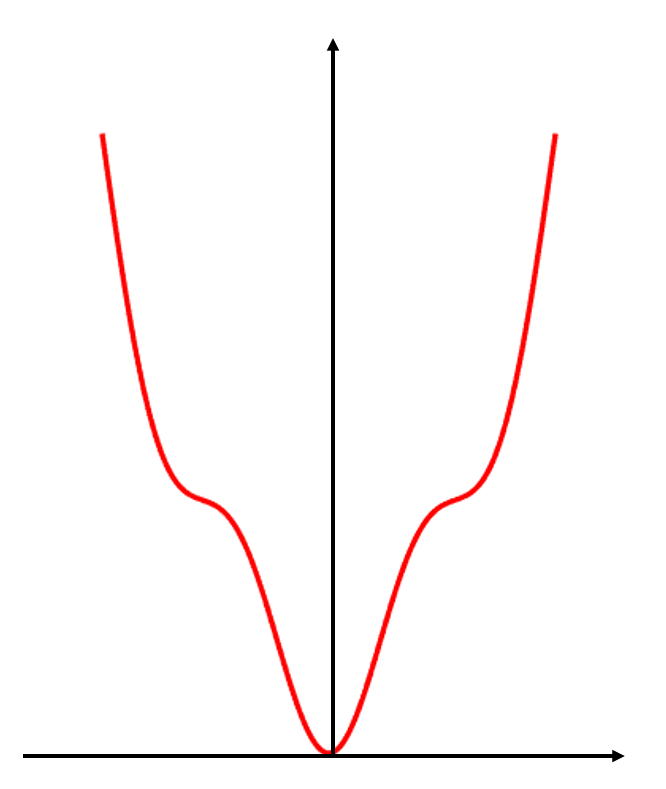}}
    \caption{The demonstration of nonconvex and invex functions. The invex function is a special class of nonconvex functions, where all stationary points are globally optimal.}
    \label{Figure:InvexityIllustration}
\end{figure}

A conventional approach to mitigating these challenges involves convex relaxation, which transforms the problem into a convex form. However, this method has several limitations. Firstly, the inherent nonconvexity of the objective function originates from the periodic nature of the cosine similarity term. Approximating it with a convex function requires specifying the period or, equivalently, imposing constraints on the norms of the signals involved. For example, \cite{gu2024robust} decompose the cosine similarity into two penalties: $2\lambda_1\bbeta_j\trans\bbeta_k-\alpha_2\|\bbeta_j\|^2$ under the assumption that $\|\bbeta_k\|_2=1$. However, due to the unknown variance of noise, it is theoretically unfeasible to determine the magnitude of the signal accurately, making the problem ill-defined when attempting convex relaxation. Secondly, as noted in \cite{duan2013differential}, the constraints derived from convex relaxation are not equivalent to the original constraints \citep{gu2024robust}. This discrepancy leads to solutions that may substantially deviate from those of the original problem. 

To address these limitations, we introduce a novel relaxation approach based on invexity, a generalization of convexity that preserves the structure of the original problem more faithfully. We prove that the optimal solution of the relaxed problem directly corresponds to that of the original problem (see Proposition \ref{prop:minimum}). Since invexity ensures that any KKT point is also a global optimum, our relaxation approach facilitates the global convergence of iterative algorithms. This characteristic not only maintains the fidelity of the original problem but also establishes a robust theoretical foundation for gradient-based algorithm development.

Now, we provide the invex relaxation of this original problem \eqref{Formulation:OriginalProblem}. We introduce notations, for $j=1,\ldots,m,i=1,\ldots,n_j$, as follows: 
\begin{equation}
\bS_{i,j}=
\begin{bmatrix}
\x_{i,j}\\-y_{i,j}
\end{bmatrix}
\begin{bmatrix}
\x_{i,j}\trans & -y_{i,j}
\end{bmatrix}=
\begin{bmatrix}
\x_{i,j}\x_{i,j}\trans & -y_{i,j}\x_{i,j}\\
-y_{i,j}\x_{i,j}\trans & y_{i,j}^2
\end{bmatrix} \in\mR^{(p+1)\times(p+1)}.\label{eq:Sij}    
\end{equation}
And the invex relaxation to the original problem is given as follows: 
\begin{equation}
\begin{split}
& \min_{\A}H_\lam(\A) , {\rm where}\quad H_\lam(\A) \defby \sum_{j=1}^m \Big\{f_j(\A_j)+ \frac{\lam}{2m}\sum_{k=1}^m g(\A_j,\A_k)\Big\}
\\
& \text{such that}\ 
\begin{array}{lc}
&(\A_j\trans\A_j)_{p+1,p+1}=1, \quad \mI_1\A_j\trans\A_j=0, j=1,\dots,m,
\end{array}
\end{split}
\label{Formulation:ParaInvexProblem}
\end{equation}
where $\A=[\A_1,\ldots,\A_m]\in \mR^{(p+1)\times m(p+1)}$ with $\A_j\in\mR^{(p+1)\times(p+1)}$, $(p+1)$-dimensional diagonal matrix $\mI_1=\diag(1,\cdots,1,0)$, 
and 
\begin{eqnarray}
&& f_j(\A_j)=\sum_{i=1}^{n_j}\tr ( \bS_{i,j}\trans\A_j\A_j\trans)/n_j, \label{function:LWi}\\
&& g(\A_j,\A_k) = -\{\tr(\A_j\A_j\trans)-1\}^{-1}\{\tr(\A_k\A_k\trans)-1\}^{-1}\tr(\A_j\A_j\trans\mI_1\A_k\A_k\trans\mI_1). \label{function: PAij}
\end{eqnarray}
The constraint $\mI_1\A_j\trans\A_j=0$ implies that $\A_j$ is a rank-1 matrix with only the $(p+1)$-th column being nonzero. Combined with $(\A_j\trans\A_j)_{p+1,p+1}=1$, this leads to the formulation of feasible $\A_j$s as: 

\begin{equation}
\A_j=\begin{bmatrix}
{\bbeta_j}\\
1
\end{bmatrix}
\begin{bmatrix}
0 &\cdots & 0 & 1
\end{bmatrix}, \ j=1,\ldots,m.
\label{invex: matrix form}
\end{equation}
Substituting $\A_j$s into $f_j(\A_j)$ and $g(\A_j,\A_k)$ recovers ${L}_j(\bbeta_j)$ and $\cos^2\inprod{\bbeta_j,\bbeta_k}$ from the original optimization problem. Proposition \ref{prop:minimum} establishes a one-to-one correspondence between the global optima of the original and relaxed problems, justifying the use of the relaxed problem as a suitable surrogate.

 {\prop Let $\hat{\bbeta}$ be any global minimum of the original problem \eqref{Formulation:OriginalProblem}. Then the induced matrix $\hat{\A}_j=\begin{bmatrix}
{\hat{\bbeta}_j}\\
1
\end{bmatrix}
\begin{bmatrix}
0 &\cdots & 0 & 1
\end{bmatrix}$ is a global minimum of the relaxed problem \eqref{Formulation:ParaInvexProblem}. Conversely, if $\hat{\A}_j$ is any global minimum of the relaxed problem \eqref{Formulation:ParaInvexProblem}, then the vector $\hat{\bbeta}_j=(\hat{\A}_j)_{1:p,p+1}$ is a global minimum of the original problem \eqref{Formulation:OriginalProblem}. 
\label{prop:minimum}       
}

Although the relaxed problem \eqref{Formulation:ParaInvexProblem} remains nonconvex, it belongs to the class of ``invex" problems, where invexity implies that all stationary points are globally optimal, as Figure \ref{Figure:InvexityIllustration} shows. Furthermore, since the Polyak-Lojasiewicz (PL) inequality, which guarantees the linear convergence rate, is related to invexity \citep{karimi2016linear}, we can expect gradient-based algorithms for this problem to achieve global convergence.

To clarify, we rigorously provide the definition of an invex function: 

{\defi[Invex function] Let $\phi(t)$ be a function defined on a set $\mathcal{T}$. Let $\eta$ be a vector-valued function defined in $\mathcal{T} \times \mathcal{T}$ such that $\eta(t_1,t_2)\trans\nabla\phi(t_2)$ is well defined $\forall  t_1,t_2\in \mathcal{T}$. Then $\phi(t)$ is a $\eta$-invex function if $\phi(t_1)-\phi(t_2)\ge\eta(t_1,t_2)\trans\nabla\phi(t_2)$, $\forall  t_1,t_2\in \mathcal{T}$.
	\label{Def:Invexity}
	}
	
\noindent	Convex functions are $\eta$-invex with $\eta(t_1,t_2)=t_1 - t_2$, demonstrating that the invexity generalizes the convexity of functions. If both the objective function and constraints are $\eta$-invex concerning the same function $\eta$, then KKT conditions are sufficient for the global optimality \citep{hanson1981sufficiency,barik2022sparse}. 

Proposition \ref{prop:InvexityForA} shows that the relaxed optimization problem (\ref{Formulation:ParaInvexProblem}) is indeed $\eta$-invex for a particular $\eta$ defined in $\mathcal{T}\times\mathcal{T}$ where $\mathcal{T}=\{\A\mid(\A_j\A_j\trans)_{p+1,p+1}=1,\mI_1\A_j\trans\A_j=0,j=1,\ldots,m\}$.

{\prop For $\A\in \mathcal{T}$, the objective function of problem \eqref{Formulation:ParaInvexProblem} is $\eta$-invex about $\A$ with $\bfeta (\A,\tilde\A) = \diag\left\{\bfeta_1(\A,\tilde{\A}),...,\bfeta_m(\A,\tilde{\A})\right\}$, where
	\begin{equation*}
	\bfeta_j\left(\A,\tilde{\A}\right)=-\tau_j(\A, \tilde{\A})\tilde{\A}_j-\tilde{\A}_j/2,
	\end{equation*}
	and
	\begin{equation}
	\tau_j(\A, \tilde{\A})= (-1/2)\left\{ \tr(\A_j\A_j\trans) - 1\right\}
    \Big\{\sum_{k\ne j}g(\tilde\A_j,\tilde\A_k)\Big\}^{-1}\sum_{k\ne j}g(\A_j,\A_k).
	\end{equation}
	\label{prop:InvexityForA}
}

The invexity of the relaxed problem ensures that every KKT point is also a global minimum. From a computational perspective, solving this problem is similar to solving a convex problem: once the KKT conditions are satisfied, the solution of the algorithm is assured to be a global minimum. This property simplifies the optimization process and guarantees the quality of the solution.

It is important to note that the original problem \eqref{Formulation:OriginalProblem} is not invex and contains numerous local minima, which can trap optimization algorithms and hinder their convergence to the global optimum (see Table \ref{tab:initialization} in Appendix \ref{app:suppResult}). Informally, the core idea behind the relaxation is to lift the problem into a higher-dimensional space through dimensionality expansion. This transformation alters the shape of the objective function around local minima, effectively eliminating these traps in the higher-dimensional space.  As a result, the relaxed problem becomes easier to solve, as it no longer suffers from the pitfalls of local minima that plagued the original formulation.

\subsection{Distributed algorithm and implementation}
Next, we develop a communication-efficient distributed algorithm to solve the relaxed problem \eqref{Formulation:ParaInvexProblem}. We solve the optimization problem \eqref{Formulation:ParaInvexProblem} through distributed projected gradient descent (PGD) with a communication cost of order $O(mp)$. 

\noindent{\bf Projected Gradient Descent}. The standard PGD update equation for our problem \eqref{Formulation:ParaInvexProblem} is given by:
\begin{equation}
\A^{(t+1)} = \mathcal{P}_{\mathcal{T}}\left[\A^{(t)} - \alpha \nabla H(\A^{(t)})\right],
\label{eq:PGD}
\end{equation}
where $H(\A)=\sum_{j=1}^m \left[f_j(\A_j)- (2m)^{-1}\lam\sum_{k=1}^m g(\A_j,\A_k)\right]$, and $\nabla H(\A)$ is its gradient. Here, $\A^{(t)}=[\A_1^{(t)},\ldots,\A_m^{(t)}]$ represents the updated $\A$ in the $t$-th iteration,  
$\alpha$ is the step size, and $\mathcal{P}_{\mathcal{T}}(\cdot)$ is the projection operator onto the feasible region $\mathcal{T}$  defined by the constraints of the problem \eqref{Formulation:ParaInvexProblem}. Since $\mathcal{T}$ is convex, the projection has a unique solution. 

Given that each $\A_j$ in the feasible region has all columns zero except the $(p+1)$-th 
column and $(\A_j)_{p+1,p+1}=1$, the gradient of $H(\A)$ with respect to $\A_j$ also has this property, except the $(p+1,p+1)$-th element differs from
 1. Therefore, the projection operator only needs to adjust the $(p+1,p+1)$-th element of the matrix $\A^{(t)}-\alpha\nabla H(\A)$ to ensure that it equals 1.

\noindent{\bf Distributed Implementation}. Since $H(\A)$ is the sum of $m$ sub-functions $H_j(\A)=f_j(\A_j)- (2m)^{-1}\lam\sum_{k=1}^m g(\A_j,\A_k)$, the PGD update \eqref{eq:PGD} can be implemented in a distributed manner. Each node $j$ computes its gradient $\nabla_{\A_j}H_j(\A)$ locally and performs one step of the PGD update independently.

Although $H_j(\A)$ depends on all $\A_j$s, the gradient $\nabla_{\A_j}H_j(\A)$ relies only on the sum of normalized $\A_j$s. With  $\bpsi = \sum_{j=1}^m (\A_j)_{1:p,p+1}/\|(\A_j)_{1:p,p+1}\| $,  we have: 
\begin{equation}
\nabla_{\A_j}H_j(\A)=2\left[\sum_{i}\bS_{i,j}\A_j/n_j-\bm\bPsi_j\A_j/\tau_j+\zeta_j\A_j/\tau_j^2\right],
\end{equation}
where $\bpsi_j = \bpsi -  (\A_j)_{1:p,p+1}/\|(\A_j)_{1:p,p+1}\| $, $\bm\bPsi_j=[\bpsi_j\trans,1]\trans[\bpsi_j\trans,1]$, $\tau_j =\tr(\mI_2\A_j\A_j\trans\mI_2)$, $\zeta_j=\tr(\mI_2\A_j\A_j\trans\mI_2\bPsi_j)$,  and $(p+1)$-dimensional diagonal matrix $\mI_2=\diag(0,\dots,0,1)$. Thus, once $\bpsi$ is known, which can be obtained from a central server, each node $j$ can compute both $\bpsi_j$ and the gradient $\nabla_{\A_j}H_j(\A)$ locally. This allows the central server to send only $\bpsi$ to each node, significantly reducing communication costs.

% To further reduce the number of communication rounds, each node performs multiple GD updates and then sends its estimates to the server. This strategy has been widely adopted in most popular federated learning algorithms, such as FedAvg \citep{mcmahan2017communication}. 

% To account for realistic network conditions where nodes may experience intermittent connectivity, it is common that some nodes are disconnected from the network because of various factors. To mimic this scenario, we adopt a randomized-block strategy where only a subset of nodes $\mathcal{A}_t\subseteq\{1,\ldots,m\}$ performs local updates and then sends their estimates to the server. 

\begin{algorithm}[!htbp]  
    \caption{Distributed reconstruction with Invex Relaxation (DIR)}  
    \begin{algorithmic}  
    \Require  
    number of nodes $m$; number of communication rounds $T$; 
    % number of local gradient epochs $T_i$, $i\in\{1,\dots,m\}$; 
    step size $\alpha$; 
    regularization parameter $\lam$;
    number of local gradient epochs $K_j$, $i\in\{1,\dots,m\}$.

    Initialize each node $j\in\{1,...,m\}$ with $\bbeta_1^{(0)}=\dots=\bbeta_m^{(0)}$. 

    Initialize the server with transmission parameter $\bpsi^{(0)}=\sum_{j=1}^{m} \bbeta_j^{(0)}/\norm{\bbeta_j^{(0)}}$.

    \Ensure $\bbeta_j^{(T)}$, $j=1,...,m.$

    \For{$t=0$ to $T-1$}
    \begin{enumerate}
        \renewcommand{\labelenumi}{\theenumi :}
        % \item (Active nodes) Server randomly selects a subset of nodes $\calA_t\subset\{1,\dots,m\}$.
        \item (Broadcast) Server sends $\bpsi^{(t)}$ to each device $j$.
        \item (Local update) For each node $j$: 
        
            $\A_j^{(0)} = [(\bbeta_j^{(t)})\trans, 1]\trans[0,\cdots, 0, 1].$

            $\bpsi_j=\bpsi^{(t)} - \bbeta_j^{(t)}/\norm{\bbeta_j^{(t)}}\quad$ $\bPsi_j =[\bpsi_j\trans,1]\trans [\bpsi_j\trans,1].$

            $\tau_j^{(k)} =\tr(\mI_2\A_j^{(k)}(\A_j^{(k)})\trans\mI_2),\quad$
            $\zeta_j^{(k)}=\tr(\mI_2\A_j^{(k)}(\A_j^{(k)})\trans\mI_2\W_j).$
            
            $\bS_{i,j}=\begin{bmatrix}\x_{i,j}\trans & -y_{i,j}\end{bmatrix}\trans\begin{bmatrix}\x_{i,j}\trans & -y_{i,j}\end{bmatrix}.$

            $\quad${\bf for} $k=0,...,K_j-1$ {\bf do} 
            \begin{equation}
                \quad\A_j^{(k+1)}=\A_j^{(k)}-\alpha \left(\sum_{i}\bS_{i,j}\A_j^{(k)}/n_j-\bm\bPsi_j\A_j^{(k)}/\tau_j^{(k)}+\zeta_j^{(k)}\A_j^{(k)}/(\tau_j^{(k)})^2\right).
            \end{equation}
            $\quad${\bf end for} 

            $\bbeta_j^{(t+1)} = \mI_2\A_j^{(K_j)}$.
        \item (Communication) Each node $j$ sends $\bbeta_j^{(t+1)}$ back to the server.
        \item (Server update) $\bpsi^{(t+1)}=\sum_{j=1}^{m} \bbeta_j^{(t+1)}/\norm{\bbeta_j^{(t+1)}}$.
    \end{enumerate}
    \EndFor 
    \end{algorithmic}
    \label{Algorithm:1}
\end{algorithm}

The overall algorithm is summarized in Algorithm \ref{Algorithm:1}. Figure \ref{Figure:converge} in Appendix \ref{app:suppResult}  illustrates the evolution of the objective function value and absolute cosine between the estimated and true signals during communication rounds for a random replication. The objective function value shows a decreasing trend as the algorithm progresses, indicating successful minimization. Meanwhile, the absolute cosine increases, reflecting improved alignment between estimated and true signals. Both metrics stabilize after around 200 rounds, suggesting efficient convergence towards optimal solutions with good reconstruction quality. Beyond a certain point, further improvements become marginal, highlighting the algorithm's efficiency.

\noindent{\bf{Regularization Parameter Selection.}} 
The proposed approach involves a regularization parameter $\lambda$ that balances the trade-off between the fidelity of the data and the level of similarity between the signals. To reduce computational cost, we adopt a warm-start strategy \citep{yu2024clustered} instead of traditional cross-validation.  
Specifically, we start with an ordered sequence of regularization parameters $\{\lambda_1,\ldots,\lambda_S\}$ $(\lambda_1<\ldots<\lambda_S)$. Then, we run the Distributed Reconstruction with Invex relaxation (DIR) algorithm with the smallest regularization parameter $\lambda_1$. Subsequently, we incrementally increase $\lambda$ to the next value when the changes in the prediction accuracy of the validation set are smaller than $10^{-4}$. The solution obtained from training with $\lambda_{s-1}$ serves as initial values for training with $\lambda_s$. 

Throughout this process, we maintain the model with the best performance in the validation set.
As our experiments show (see Figure \ref{Figure:Warming-up} in Appendix  \ref{app:suppResult}), the validation accuracy usually improves with increasing $\lambda$, especially for smaller $\lambda$.
However, excessive regularization, especially for weakly similar signals with $\theta_{\max}=\pi/3$, can degrade performance by over-constraining signal alignment and suppressing meaningful directional variation.

Once the optimal $\lambda^*$ is identified based on the best validation performance, we retrain the model using this regularization parameter until full convergence. Using the warm-start scheme significantly reduces the running time, for example, reducing it to only 5 seconds with warm-start compared to 49 seconds when selecting the regularization parameters by separate tuning on the synthetic dataset with low similarity ($\theta_{\max}=\pi/3$) between signals, as quantified in Table \ref{Table:TuningTime} in Appendix \ref{app:suppResult} shows.

\section{Theoretical guarantees} \label{sec:theorem}
In this section, we establish theoretical guarantees for the proposed method. Specifically, we prove the uniqueness of the global minimum of the original problem \eqref{Formulation:OriginalProblem} under mild regular conditions \ref{Theorem:0}. We then provide a nonasymptotic error bound for $\|\hbeta_j/a_j-\bbeta^*_j\|$, where $a_j$ is specified in Theorem \ref{Theorem:1}. Finally, we analyze the convergence rates of Algorithm \ref{Algorithm:1} in Theorem \ref{Theorem:2}. All the proofs are provided in Appendix \ref{app:proof}. 

For the theoretical analysis, we assume the following regular conditions: 
\begin{enumerate} 
\item[(C1)] The measurement vectors $\x_{i,j}$s are i.i.d. random vectors sampled from the multivariate normal distribution $\mathcal{N}({\bf 0},\bSig)$. 
\item[(C2)] % {(condition on \x)}
    There exist two positive constants $C_{\min}$ and $C_{\max}$ such that $0 < C_{\min} \le \gamma_{\min}(\bSig)\le \gamma_{\max}(\bSig) \le C_{\max}<\infty$, where $\gamma_{\min}(\bSig)$ and $\gamma_{\max}(\bSig)$ denote the smallest and the largest eigenvalues of $\bSig$, respectively. 
\end{enumerate}
Condition (C1) is common in 1-bit CS \citep{huang2018robust, plan2013one}. Condition (C2) assumes that the covariance matrices are positive definite. Without loss of generality, we assume $\norm{\bbeta_j}_{\bSig} = 1$ for $j=1,\cdots,m$, where $\norm{\bbeta}_{\Sigma} = \bbeta\trans\bSig\bbeta$ is the elliptic norm of $\bbeta$ with respect to $\bSig$, following the assumption from \cite{huang2018robust}. And we denote the condition number of $\bSig$ as $\kappa(\bSig)$.

We begin by establishing the uniqueness of the global minimum for the original optimization problem \eqref{Formulation:OriginalProblem}.

{\theo[Uniqueness of Global Minimum] Under Conditions (C1) and (C2), the optimization problem \eqref{Formulation:OriginalProblem} has a unique KKT point with global optimality.  
\label{Theorem:0}
}

Theorem \ref{Theorem:0} shows that, under mild conditions, the original problem possesses a single global minimum. 
To enable this, the invexity property of the relaxed problem \eqref{Formulation:ParaInvexProblem} guarantees that this solution corresponds to the unique KKT point. Crucially, Proposition \ref{prop:minimum} demonstrates that the original problem \eqref{Formulation:OriginalProblem}possesses a single global minimum, similar to the relaxed formulation.
These results ensure both the well-posedness of the problem and the reliability of the obtained solutions, which provides a solid foundation for the subsequent theoretical analysis of estimation error bounds and algorithmic convergence.

\subsection{Nonasymptotic error bound of estimator $\hat\bbeta$}
In this subsection, we provide a nonasymptotic error bound for the unique global minimum $\hbeta$ of problem \eqref{Formulation:OriginalProblem}. Let $N=\sum_{j=1}^m n_j$ denote the total sample size from $m$ nodes. 

To derive the error bound, we introduce two additional conditions:  
\begin{enumerate}
\item[(C3)] There exist constants $0<r<R\leq 1$ such that $rN\leq n_j \leq RN$ for all $j=1,\ldots,m$.
\item[(C4)] % {(condition on \x)}
There exists a constant $0\le\theta< \pi/2$ such that $|\cos\inprod{\bbeta_j,\bbeta_k}|\geq \cos\theta$ for any $j,k=1,\ldots,m$. 
\end{enumerate}
Condition (C3) ensures that the sample size at each node is neither disproportionately large nor small compared to the total sample size.
Condition (C4) imposes a lower bound on the cosine similarity between any pair of true signals $\bbeta_j$ and $\bbeta_k$, reflecting a certain level of alignment among the signals.  

%todo order 版本, 放在正文
{\theo [Nonasmptotic Error Bound] Under Conditions (C1)-(C4), if $n_j > O(p\log{p} \vee mr^{1/2})$ and $\cos\theta > O\{(p\log{p}/n_j)^{1/2}\}$ for $j=1,\dots,m$, then it holds that
 \begin{equation}
        \begin{split}
            \norm{\hbeta_j/a_j - \bbeta_j^*} \le&
            O\left\{(1-\lam)(p\log{p}/n_j)^{1/2} + \lambda(p\log{p}/N)^{1/2}
            +\lambda p^{1/2}/n_j\right\},
        \end{split}
        \label{Theorem1:Result1}
\end{equation}
with probability at least  $1-O\left(m\exp(-C_1C_2^2p)+m/p^3\right)$, where $a_j=(2q_j-1)\{\pi(\sigma_j^2+1)/2\}^{-1/2}$. Here $C_1$ and $C_2$ are some generic constants independent of $n_j$, $p$ or $m$.

Furthermore, when $\alpha_0(m,r,\theta)^{-1}(1-r^{1/2}) \le \lambda \le \alpha_0(m,r,\theta)^{-1}(1+r^{1/2})$, with the same probability, we have
\begin{equation}
    \norm{\hbeta_j/a_j - \bbeta_j^*} \le
    O\left\{ (p\log{p}/N)^{1/2} + p^{1/2}/n_j\right\}.
    \label{Theorem1:Result2}
\end{equation}
Here $\alpha_0(m,r,\theta)$ is a factor of $m$, $r$ and $\theta$, but independent of $n_j$, $p$ or $m$.
\label{Theorem:1}
}

% two main results
Theorem \ref{Theorem:1} provides a nonasymptotic error bound for $\|\hbeta_j/a_j - \bbeta_j^*\|$,
where the convergence rate of $\hbeta_j$ in the result \eqref{Theorem1:Result1} is minimax optimal up to a logarithmic factor in $\ell_2$-norm in parameter space $\mR^p$ \citep[see][Chapter 15]{wainwright2019high}. The logarithmic factor is due to the loss of information with the l-bit quantization, compared to complete data quantization. As the result \eqref{Theorem1:Result2} shows, given $m$ of order at least $O(\log{p} )$ and $\lam$ in the appropriate range, the convergence rate of $\hbeta_j$ is minimax optimal up to a constant factor in $\ell_2$-norm in parameter space $\mR^p$. The specific factor of these results can be found in Lemma \ref{lem:Specific_Error} in Appendix \ref{app:proof}, where a wider and more precise range for regularization parameter, $\lam_L\le\lam\le\lam_U$, is given.

Theorem \ref{Theorem:1} presents two key results. 
The first result \eqref{Theorem1:Result1} decomposes the estimation error bound into three components: local quantization errors, network-wide quantization errors, and penalty function errors, with the orders $O\{(1-\lam)(p\log{p}/n_j)^{1/2}\}$, $O\{\lam(p\log{p}/N)^{1/2}\}$, and $O(\lam p^{1/2}/n_j)$ respectively. These terms capture the effects of noise and nonlinearity introduced by 1-bit quantization at individual nodes and across the network, as well as the impact of the penalty function used in the formulation. The specific factors of these terms increase with higher noise variance $\sigma_j^2$, condition number $\kappa(\bSig)$, and sign-flip probability $q_j$ closer to 0 or 1.
As the specific results in Lemma \ref{lem:Specific_Error} show, the local and network-wide quantization errors have a $O(\log{p})$ information loss caused by the 1-bit measurements, compared to the minimax optimal rate. 
However, the factor of penalty function errors $O\left( p^{1/2}/n_j\right)$ remains constant and decreases with greater cosine similarity $|\cos\theta|$ between signals.
When $\lambda=0$, the estimation error can exactly recover the standard LS solution, which estimates each signal separately using LS methods. In this case, the error bound is $ O\{(p\log{p}/n_j)^{1/2}\}$ \citep{huang2018robust}.  

The second result \eqref{Theorem1:Result2} demonstrates that with an appropriate choice of the regularization parameter $\lambda$ within a defined range,
% ($\lambda_{L}\leq \lambda \leq \lambda_{U}$)
our estimate can achieve a faster statistical convergence rate than the standard LS solution. Specifically, the overall convergence rate is ($ O\{(p\log{p}/N)^{1/2}+p^{1/2}/n_j\}$), which simplifies to $ O\{(p\log{p}/N)^{1/2}\}$ when $m \ll n_j$, or $ O(p^{1/2}/n_j)$ otherwise,  when $\lambda_{L}\leq \lambda \leq \lambda_{U}$ as given in Lemma \ref{lem:Specific_Error}. 
When the number of similar nodes satisfies $m$ is of order at least $O(\log{p})$ and $\lam$ is appropriately chosen, the information sharing among nodes effectively compensates for the quantization error introduced by 1-bit CS measurements, enabling our method to achieve the minimax optimal rate.

The error bound in \eqref{Theorem1:Result1} reveals an important trade-off governed by the regularization parameter $\lam$. As $\lambda$ increases, our method reduces the influence of local quantization errors by incorporating information from other nodes, accelerating convergence. However, this benefit comes at a cost: excessive $\lambda$ can amplify noise and nonlinearity effects, hence necessitating the upper bound limit $\lambda_{U}$. The optimal range for $\lambda$ varies based on factors such as noise variance $\sigma_j^2$, sign-flip probability $q_j$, condition number $\kappa(\bSig)$, and cosine similarity $|\cos\theta|$. 
Both $\lam_L$ and $\lam_U$ increase with higher cosine similarity, 
indicating that more inter-node information should be employed in high similarity scenarios.
Conversely, these bounds decrease with larger noise variance, indicating that less information sharing is preferable in noisy environments.

To achieve the theoretical bounds in Theorem \ref{Theorem:1}, certain requirements must be satisfied regarding local sample size $n_j$ and cosine similarity $\cos\theta$. The required sample size and similarity become stricter under conditions of high noise and nonlinearity. When the sample size $n_j$ is large enough, the similarity requirement becomes less stringent, permitting $\cos\theta$ to approach zero. The adaptive nature of our method allows it to identify and leverage the degree of similarity between signals, meaning that even if the true signals are not perfectly aligned, the method can still perform effectively. Such flexibility is crucial in practical applications where perfect alignment might not always be feasible or realistic.

\subsubsection{The perspective of corrected LS}

\begin{figure}[!htbp]
    % \flushleft %左对齐
    \centering
        \subfigure[A similar case]{\label{Figure:CorrectionSimilar}
        \includegraphics[width =0.4\textwidth]{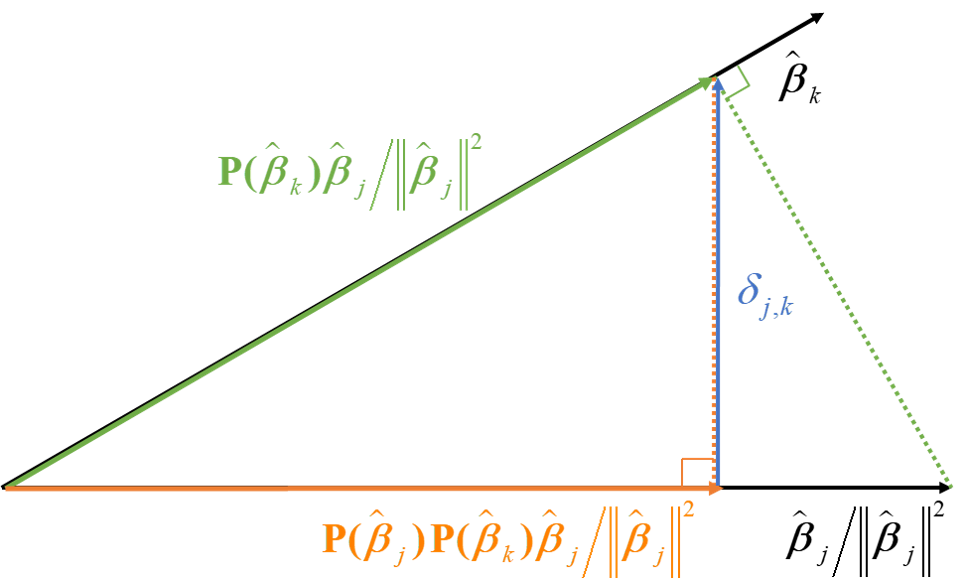}}
        \quad\quad
        \subfigure[A dissimilar case]{\label{Figure:CorrectionUnsimilar}
        \includegraphics[width =0.4\textwidth]{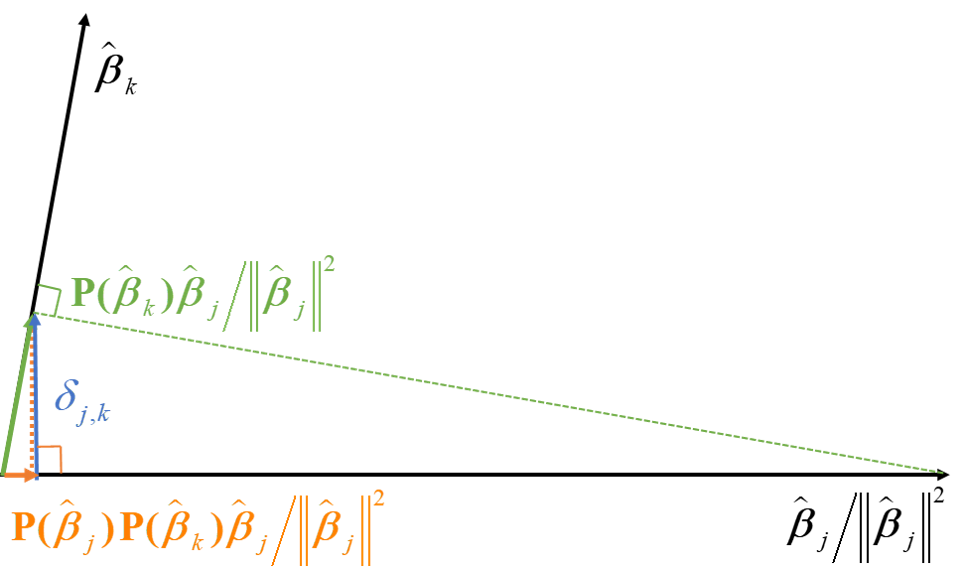}}
    \caption{ Illustration that correction $\bdelta_{j,k}$ contributed by each node is determined by inter-node similarity and signal intensity $\norm{\hbeta_j}$. The correction $\bdelta_{j,k}$ has a higher impact $\norm{\bdelta_{j,k}}$ with higher inter-node similarity and lower local signal intensity.}
    \label{Figure:Addptive}
\end{figure}

The enhanced convergence rate of our proposed method results from its ability to adaptively introduce similar-node information. Our method achieves adaptivity to heterogeneity through two distinct mechanisms: inter-node similarity and SNR of the signals to reconstruct. First, the squared cosine similarity penalty adaptively incorporates highly similar node information and mitigates the influence of heterogeneous nodes.
Second, for nodes significantly affected by noise and sign flips, additional inter-node information is imported to enhance estimation accuracy. Specifically, our method serves as a bias correction mechanism for the local LS solution:  
\begin{align}
        \hbeta_j&=(\X_{j}\trans\X_{j})^{-1}\X_{j}\trans\y_{j}+(\X_{j}\trans\X_{j})^{-1}\bdelta_j,
        \quad
        \bdelta_j = \sum_{k\ne j } \bdelta_{j,k} 
    \label{eq:BiasCorrection}
    \\ 
    \bdelta_j &= ({\lambda}/{m})\left\{ \I_p- \P(\hbeta_j)\right\}\P(\hbeta_k)\hbeta_j/\norm{\hbeta_j}^2,
    \label{eq:deltak}
\end{align}
% $\P(\B) = \B(\B\trans\B)^{-1}\B\trans$
where $\I_p$ is a $p+1$-dimensional identity matrix and $\P(\bbeta) = \bbeta\bbeta\trans/\norm{\bbeta}^{2}$ is the projection matrix of vector $\bbeta$.
Equation \eqref{eq:BiasCorrection} shows that the estimate $\hbeta_j$ at node $j$ is the sum of the LS estimate and a correction term $\bdelta$. 

The correction term, as detailed in \eqref{eq:deltak}, is determined by the regularization parameter $\lambda$, the projections of the signal estimates from other nodes onto the local estimate, and the SNR characteristic $\norm{\hbeta_j}$. As Figure \ref{Figure:Addptive} illustrates, the corrections from high-similarity nodes contribute more significantly to $\bdelta_j$ with higher-level impact $\norm{\bdelta_{j,k}}$, while those with lower similarity have less correction impact $\norm{\bdelta_{j,k}}$.

Besides, the overall correction magnitude $\norm{\bdelta_j}$ of node $j$ exhibits an inverse relationship with $\norm{\hbeta_j}$, where this estimated signal magnitude reflects the SNR through $\norm{a_j\bbeta^*_j}$. As Lemma \ref{Lemma: ls solution 1} in Appendix \ref{app:TechnicalLemma} and Theorem \ref{Theorem:1} show, given the original signal intensity $\norm{\bbeta_j^*}$, the estimated signal magnitude $\norm{\hbeta_j}$ decreases with smaller $|a_j|$, under conditions of higher noise variance or increased sign-flip probability.

Overall, this similarity-based adaptive information contribution ensures that the method can dynamically adjust based on the degree of similarity between signals, enhancing accuracy and reliability in distributed learning scenarios.

\subsection{Convergence rate of Algorithm \ref{Algorithm:1}}

In this subsection, we derive the convergence rate of Algorithm \ref{Algorithm:1}. We denote $\Delta_0 = G_\lam(\beta^{(0)}) - G_\lam(\beta^{(0)})$. Let $m_1 =  a_{\min}C_{\max}^{-1/2}\kappa(\bSig)^{-1}/3$, $m_2 = 3p^{1/2}\kappa(\bSig)C_{\min}^{-1/2}a_{\max}$, where $\kappa(\bSig)$ is the condition number of $\bSig$, $a_{\min} = \min_j{a_j}$ and $a_{\max} = \max_j{a_j}$. We have the following results.

{\theo [Specific convergence rate of Algorithm \ref{Algorithm:1}]
   Assume Conditions (C1)-(C4), $n_j>O(p\log{p} \vee mr^{1/2})$, $\cos\theta\ge O\{(p\log{p}/n_j)^{1/2}\} $ for $j=1,\ldots,m$, and $\alpha_0(m,r,\theta)^{-1}(1-r^{1/2}) \le \lambda \le \alpha_0(m,r,\theta)^{-1}(1+r^{1/2})$. Set the step size $\alpha=1/M$, with $\nu=C_{\min}/2 $ and $M=3C_{\max}+24\lam(1+m_1)m_1^{-2}$. Then, with probability at least $1-O\left(m\exp(-C_1C_2^2p)+m/p^3\right)$, for any $T\geq 1\vee T_0 $, 
            \begin{equation}
        \begin{split}
            \norm{\bbeta_j^{(T)}/a_j - \bbeta_j^*} \le &
            O \left\{(2/\eta)^{1/2}(1-\nu/M)^{T/2}  
            \Delta_0\right\} +O\{ (p\log{p}/N)^{1/2} + p^{1/2}/n_j \},
        \end{split}
        \end{equation}
    where $\eta=C_{\min}/2$ and $T_0 \defby 2\log(\hat\alpha(\lambda,p)\Delta_0^{-1})/\log(1-\nu/M)$, where $\hat\alpha(\lambda,p)$ is a factor of $\lambda$ and $m_2$, given in Appendix \ref{app:proofTheorem2}. 
    Here $C_1$ and $C_2$ are some generic constants from Theorem \ref{Theorem:1}, independent of $n_j$, $p$, or $m$. 
    \label{Theorem:2}
}
    
Theorem \ref{Theorem:2} establishes the global convergence rate of our algorithm solution $\bbeta_j^{(T)}$, which is independent of initialization, as Table \ref{tab:initialization} in Appendix \ref{app:suppResult} shows. For large enough $T$, the convergence rate of $\hat\bbeta_j^{(T)}$ is $O\{ (p\log{p}/N)^{1/2} + (p^{1/2}/n_j) \}$, which matches the minimax optimal rate for estimating $\bbeta_j^*$ in $\ell_2$-norm in $\mR^p$ globally. Hence, the proposed distributed estimators converge to the global estimators. 

Comparing Theorem \ref{Theorem:1} and Theorem \ref{Theorem:2}, we see some important trade-offs in distributed 1-bit CS. First, the estimation error of $\bbeta^{(T)}$ with small $T$ is larger than the pooled version $\hat\bbeta$ because it leverages summary information rather than individual data points.
As $T$ increases, the accuracy improves, but so does the communication cost between the server and nodes. Therefore, balancing communication efficiency and estimation accuracy is crucial. The choice of $T$ should consider practical constraints such as network bandwidth, computational resources, and the desired level of estimation accuracy.

\section{Numerical simulations}\label{sec:simulation}
In this section, we conduct extensive numerical experiments to evaluate the performance of the proposed approach. First, we outline the experimental setup. Then, we compare the estimation performance of different methods. Ablation studies conclude this section. All experiments are repeated 100 times. More experimental details are provided in Appendix \ref{app:suppResult}. All the computations are performed on a sixteen-core laptop with 2.59 GHz and 32 GB RAM using MATLAB 2022a. Our code is released at \url{https://github.com/Lear24/DIR-for-CS.git}.

\subsection{Experiments settings}
\noindent{\bf Data Generation Procedure.}  
To simulate realistic conditions for our study, we generate synthetic data as follows. 
For each measurement $i$ on node $j$, the measurement vectors $\x_{i,j}$s are drawn from a multivariate normal distribution $\mathcal{N}({\bf 0,\Sigma})$, 
where the $(k,l)$-th entry of the covariance matrix ${\bf \Sigma}_{k,l}=0.3^{|k-l|}$. The noise vectors $\bm\varepsilon_{j}$ 
for each node $j$ are generated from $\mathcal{N}({\bf 0},\sigma_j^2\bf{I})$. To introduce heterogeneity in noise levels across nodes, we set the parameters for noisy measurement $(\sigma_j,q_j)$ to either(0.1, 0.75) or (0.2, 0.125) with equal probability.

The signal dimension is fixed at $p=20$.  
We provide a procedure to introduce heterogeneity in the direction of signals across nodes while ensuring a certain level of similarity or dissimilarity. Specifically, we first generate the signal $\bbeta_1$ for the first node, where its entries are independently drawn from a Bernoulli distribution $Bernoulli(1,0.5)$. For each subsequent node $j\geq 2$, we draw an angle uniformly from the interval $(0,\theta_{\max})\cup(\pi-\theta_{\max},\pi)$, where $\theta_{\max}\in(0,\frac{\pi}{2})$. Additionally, $p-2$ basis vectors are randomly selected from an orthogonal unit basis to define rotation axes. Each signal $\bbeta_j^*$ for node $j\geq 2$ is obtained by rotating $\bbeta_1^*$ around the chosen rotation axes by the specified cosine similarity value. This ensures that the cosine similarity between any two signals is greater than $\cos\theta_{max}$ or smaller than $-\cos\theta_{max}$, thereby introducing controlled variability in signal directions while maintaining a minimum level of similarity or dissimilarity.  

\noindent{\bf Comparison Methods.} We compare the performance of the proposed approach with four state-of-the-art methods: (1) separate least squares (SLS), which applies LS to each node separately, i.e., solves the original problem \eqref{Formulation:OriginalProblem} with $\lambda=0$; (2) pool least squares (PLS), which combines measurements from all nodes and applies LS to the aggregated data; 
(3) distributed robust decoding (DRD)\citep{chen2023distributed}, which assumes all nodes share the same signal and decodes the signal in a distributed manner, and  
(4) centralized invex relaxation (CIR), which solves the invex relaxed problem \eqref{Formulation:ParaInvexProblem} through the PGD algorithm in a centralized setting, where all measurements are processed at a server. Both the proposed approach and CIR adopt the warm-start strategy to select the regularization parameter $\lambda$ from the set $\{0.4,0.6,\ldots,1.6\}$. We conduct the proposed approach by performing multiple local GD updates on each node \citep{mcmahan2017communication}, to reduce the communication cost.

\noindent{\bf Evaluation Metrics.} We evaluate the estimation performance using two primary metrics. The first is the $l_2$-error between the estimated signal and true signal, which is defined as $\|\hat\bbeta_j/\|\hat\bbeta_j\|-\bbeta_j^*/\|\bbeta_j^*\|\|$ if $q_j>0.5$, and $\|\hbeta_j/\|\hbeta_j\|+\bbeta_j^*/\|\bbeta_j^*\|\|$ otherwise. A smaller $l_2$-error indicates higher accuracy in estimating $\bbeta_j$. To evaluate the overall estimation performance across all nodes, we report the average $l_2$-error, defined as 
$\sum_{j=1}^m \|\hbeta_j/\|\hbeta_j\|\pm\bbeta_j^*/\|\bbeta_j^*\|\|/m$ where the sign is chosen based on the value of $q_j$. The second is  absolute cosine, defined as ${|\hat\bbeta_j\trans\bbeta_j^*|}/{\|\hat\bbeta_j\| \|\bbeta_j^*\|}$. Values closer to 1 indicate more accurate estimates. All simulation results are based on 100 independent replications to ensure statistical reliability. 
     
\begin{figure}[!htbp]
    % \flushleft %左对齐
    \centering
    \subfigure[$\theta_{\max}=\pi/3$]{
        \includegraphics[width =0.3\textwidth]{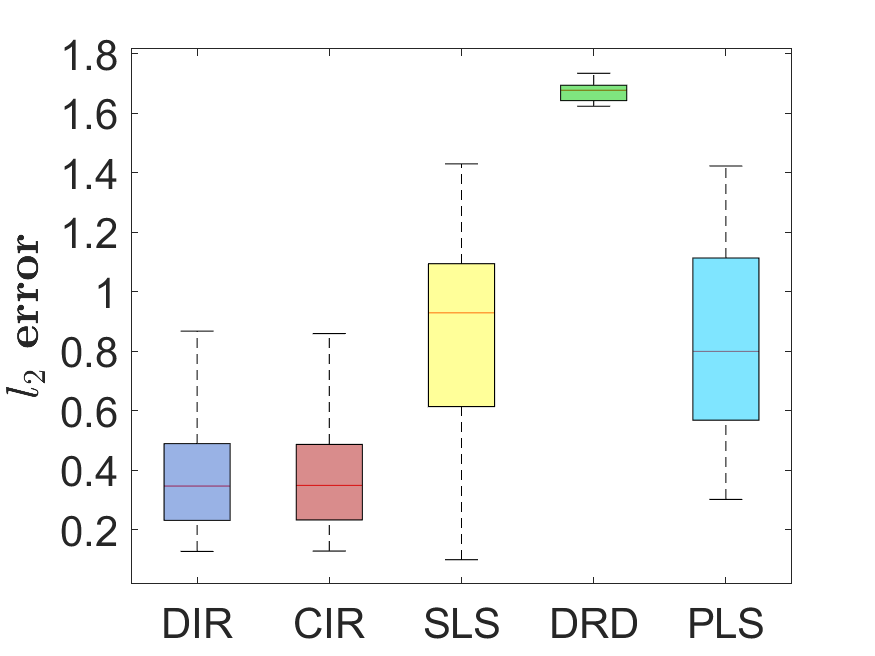}}
    \subfigure[$\theta_{\max}=\pi/4$]{
        \includegraphics[width =0.3\textwidth]{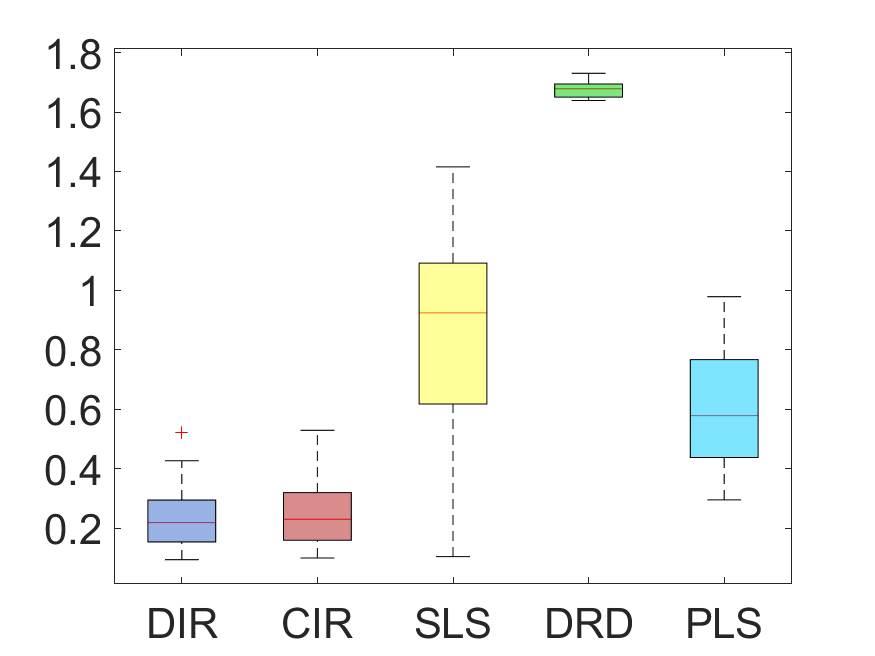}}
    \subfigure[$\theta_{\max}=\pi/8$]{
        \includegraphics[width =0.3\textwidth]{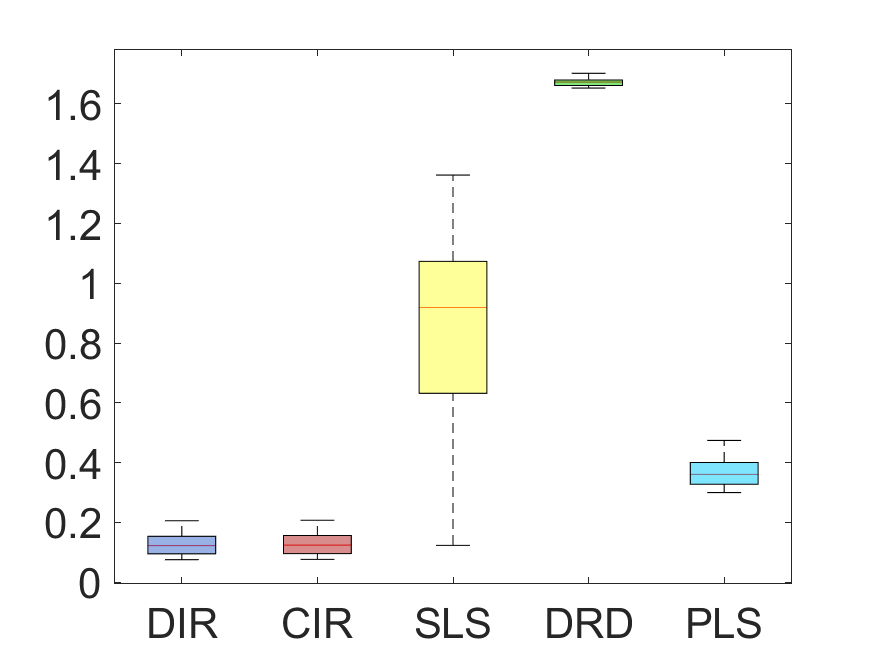}}
    \caption{Distribution of $l_2$-error across nodes for the proposed approach DIR (purple), SLS (red), PLS (yellow), DRD (green), and CIR (blue) under 
    different levels of angular similarity: $\theta_{\max}=\pi/3$ (a), $\theta_{\max}=\pi/4$ (b), and $\theta_{\max}=\pi/8$ (c). The simulations fix $N=2400$, $p=20$ and $m=30$.}
    \label{Figure: distribution across nodes}
\end{figure}

\subsection{Performance evaluation}
Figure \ref{Figure: distribution across nodes} and \ref{Figure:DistributionAcrossNodes2} depict the distributions of $l_2$-error and absolute cosine, respectively, between estimated signals and true signals across all nodes for datasets with $N=2400$, $m=30$, and $\theta_{\max}\in\{\pi/3,\pi/4,\pi/8\}$ representing low, medium, and high levels of similarity among signals. The number of measurements at each node follows a power-law distribution, as Figure \ref{Figure:DataDistribution} shows in Appendix \ref{app:suppResult}. The proposed approach consistently demonstrates superior estimation performance across the entire simulation spectrum. Its close agreement with the centralized version (CIR) validates the accuracy and effectiveness of the distributed computation. In contrast, SLS performs worse than the proposed approach, and its performance is not strongly affected by the similarity level across signals, as it estimates each signal separately. PLS exhibits good performance when the signals have a high similarity level, for example, when $\theta_{max}=\pi/8$. However, as expected, its performance deteriorates significantly as signal heterogeneity increases. DRD exhibits the poorest estimation performance due to its fundamental assumption of signal uniformity and its reliance on approximating. The main reason lies in its approximation of the Hessian matrix using only the local Hessian matrix from the first node; this approximation becomes poor when signals across nodes differ significantly. These results underscore the importance of considering inter-signal heterogeneity, highlighting that methods accounting for angular similarities among signals outperform those that ignore these similarities or incorrectly assume uniformity across nodes.

\begin{figure}[!htbp]
    % \flushleft %左对齐
    \centering
    \subfigure[$\theta_{\max}=\pi/3$]{
        \includegraphics[width =0.3\textwidth]{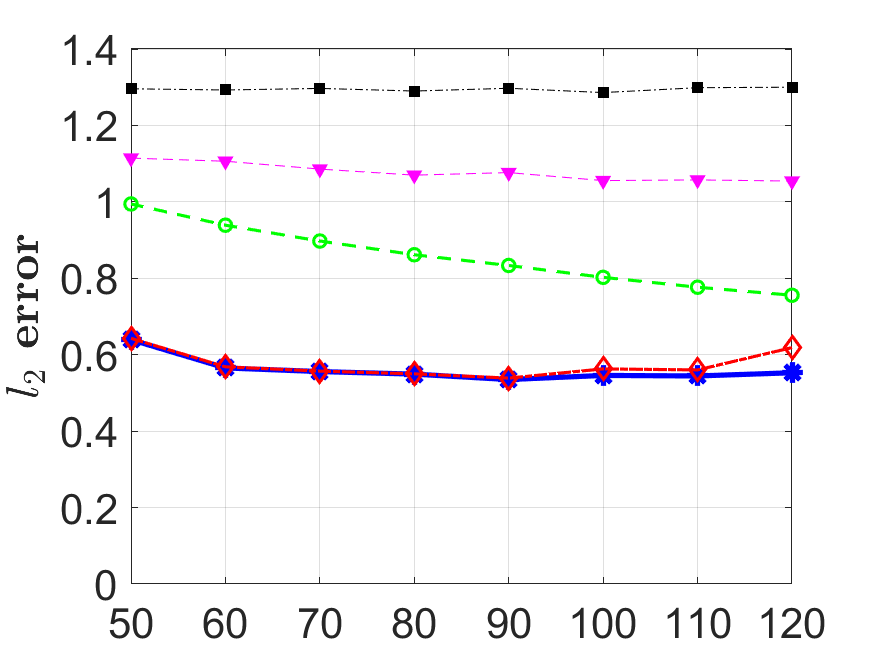}}
    \subfigure[$\theta_{\max}=\pi/4$]{
        \includegraphics[width =0.3\textwidth]{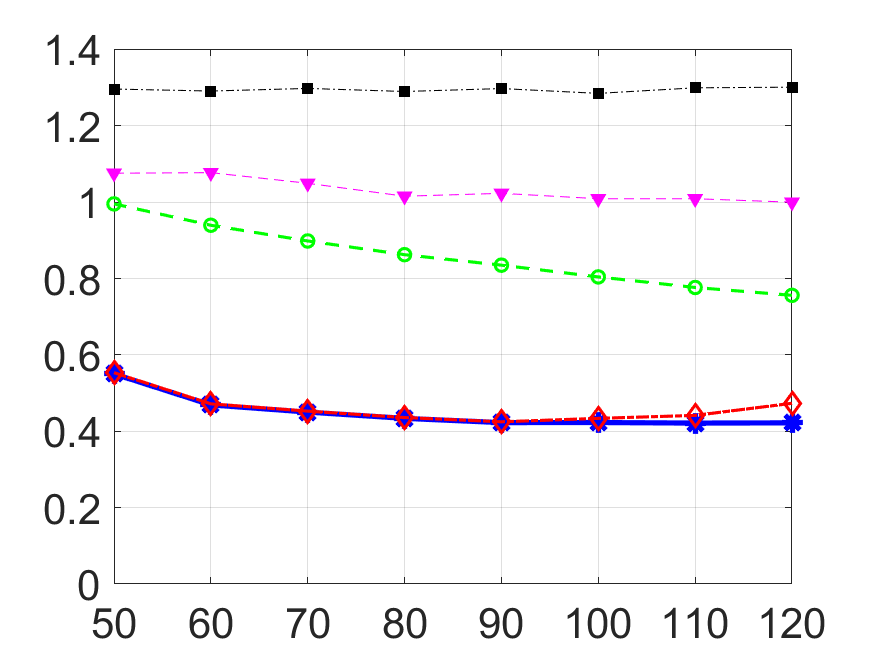}}
    \subfigure[$\theta_{\max}=\pi/8$]{
        \includegraphics[width =0.3\textwidth]{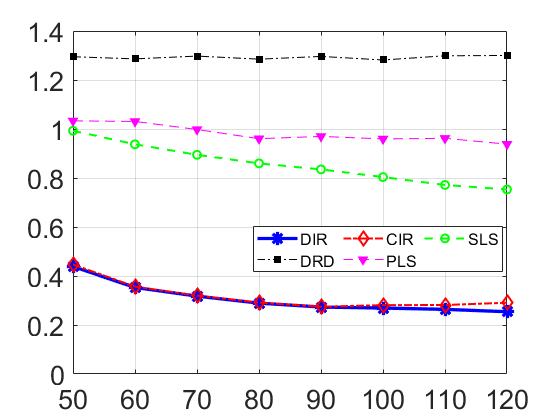}}
    \caption{Average $l_2$-error against the number of local measurements for the proposed approach DIR (solid-star), SLS (dash-circle), PLS (dash-triangle), DRD (dot-dash-square), and CIR (dot-dash-diamond) under 
    different levels of angular similarity: $\theta_{\max}=\pi/3$ (a), $\theta_{\max}=\pi/4$ (b), and $\theta_{\max}=\pi/8$ (c). The simulations fix $p=20$ and $m=30$.}
    \label{Figure: Effect of local sample size}
\end{figure}

%% fix m, increase n_i %%
To examine the effect of local measurement quantity, we conduct experiments with a fixed number of nodes $m=30$, and vary the number of local measurements $n_j\equiv n\in\{50,60, \ldots, 120\}$. The results are presented in Figure \ref{Figure: Effect of local sample size} and \ref{Figure: Effect of local sample size cos} in Appendix \ref{app:suppResult}. The proposed approach consistently outperforms other methods across the entire simulation spectrum. As $n$ increases, the average $l_2$-error of the proposed estimate decreases gradually, aligning with the findings of Theorem \ref{Theorem:1}. As expected, SLS exhibits the fastest decrease in $l_2$-error with increasing $n$, due to its independent estimation for each node. However, despite this rapid improvement, SLS remains inferior to the proposed approach in terms of overall accuracy. The proposed method's superior performance highlights its ability to leverage both individual node data and inter-node similarities effectively, enabling it to maintain superior estimation accuracy across varying local sample sizes.

%\subsection{Effect of heterogeneity}
\subsection{Ablation study}
In this subsection, we explore how the reconstruction performance of our proposed approach is affected by several key factors: the number of nodes, the distribution of local measurement sizes, noise intensity, and sign-flip probability.

\noindent{\bf Number of Nodes.} 
We investigate the effect of varying the number of nodes $m$ in $\{2^1,2^2,\ldots,2^{8}\}$ while fixing the local measurement size $n_j\equiv n=60$. The results are depicted in Figure \ref{Figure: m vary} and \ref{Figure: m vary cos} in Appendix \ref{app:suppResult}. The average $l_2$-error of the proposed estimate exhibits distinct behavior depending on the relationship between $m$ and $n$. When $m<n$, the $l_2$-error scales linearly with the number of nodes on a log-log scale. This observation aligns with the theoretical results in Theorem \ref{Theorem:1}, which states that the $l_2$-error is $O\{(p\log{p}/nm)^{1/2}\}$ with a high probability. Conversely, when $m>n$, the $l_2$-error remains nearly constant, echoing Theorem \ref{Theorem:1}'s prediction that the $l_2$-error becomes $O(p^{1/2}/n_j)$ with a high probability, independent of $m$. 

\begin{figure}[!htbp]
    % \flushleft %左对齐
    \centering
    \subfigure[$\theta_{\max}=\pi/3$]{
        \includegraphics[width =0.3\textwidth]{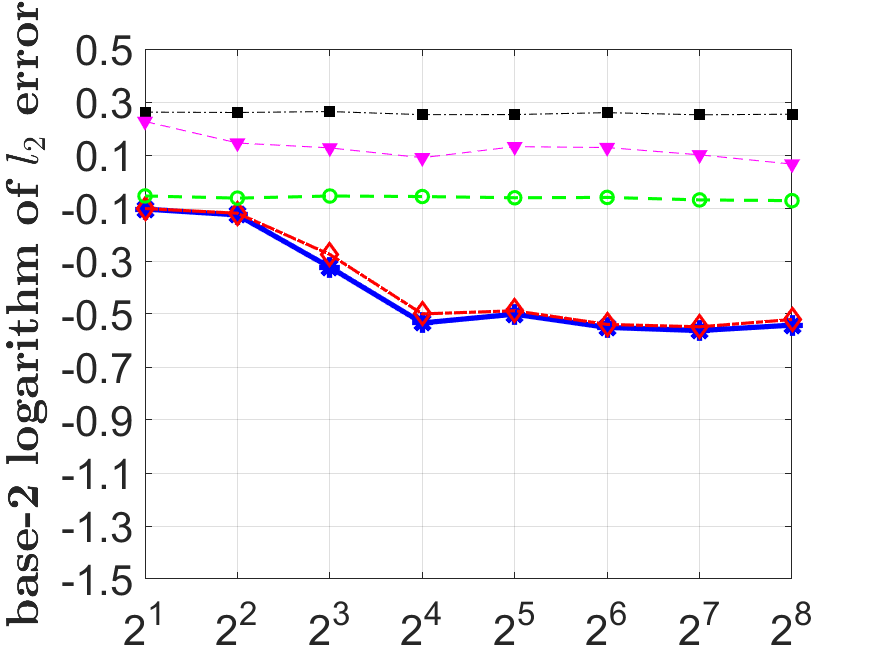}}
    \subfigure[$\theta_{\max}=\pi/4$]{
        \includegraphics[width =0.3\textwidth]{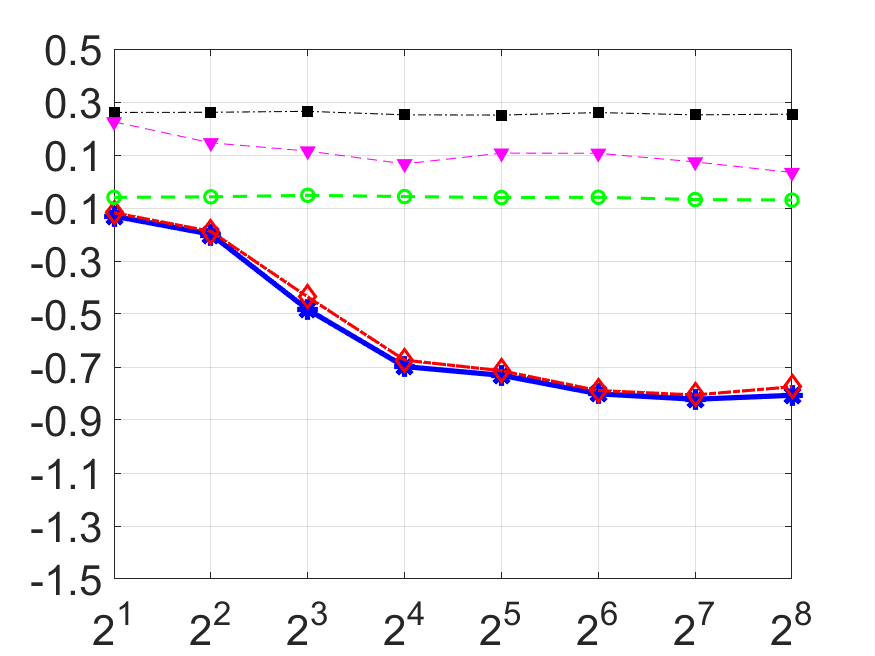}}
    \subfigure[$\theta_{\max}=\pi/8$]{
        \includegraphics[width =0.3\textwidth]{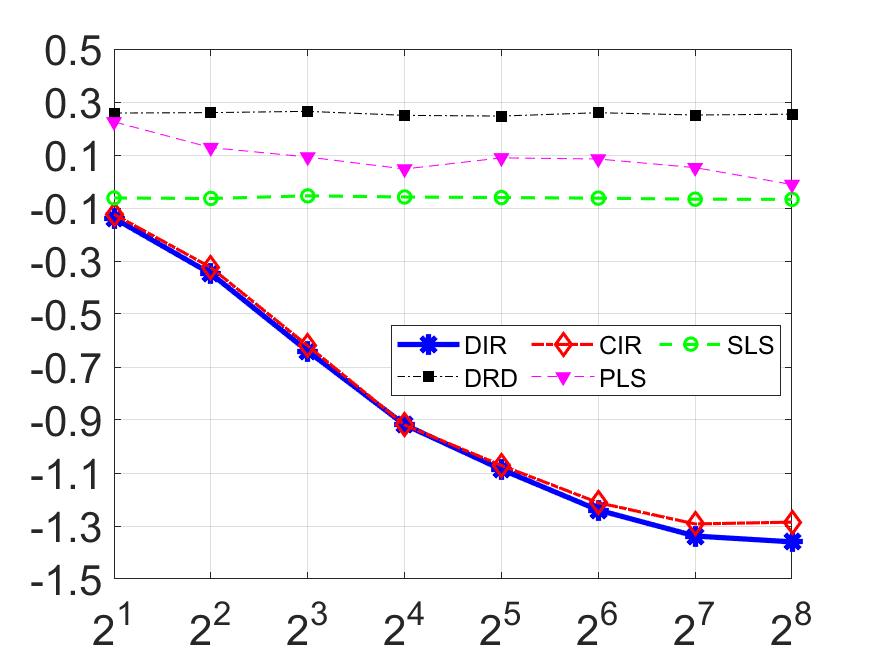}}
    \caption{Average $l_2$-error against the base-2 logarithm number of nodes for the proposed approach DIR (solid-star), SLS (dash-circle), PLS (dash-triangle), DRD (dot-dash-square), and CIR (dot-dash-diamond) under 
    different levels of angular similarity: $\theta_{\max}=\pi/3$ (a), $\theta_{\max}=\pi/4$ (b), and $\theta_{\max}=\pi/8$ (c). The simulations fix $p=20$ and $n=60$.}
    \label{Figure: m vary}
\end{figure}

\noindent{\bf Distribution of Local Measurement Size.}
To examine the effect of the distribution of the local measurement sizes, we simulate scenarios where the number of measurements follows a Dirichlet($\alpha$) distribution with $\alpha\in\{0.3,0.4,0.5,0.6,0.7\}$. A smaller $\alpha$ indicates more pronounced heterogeneity in the number of measurements between nodes. This setup mimics real-world conditions where different devices or nodes may have varying levels of data collection capabilities. The supporting results are included in Appendix \ref{app:suppResult}. As shown in Figure \ref{Figure: Effect of local sample size in the Dirichlet} and \ref{Figure: Effect of local sample size in the Dirichlet cos}, our proposed approach maintains robustness against these measurement count imbalances. 

% \begin{figure}[!htbp]
%     % \flushleft %左对齐
%     \centering
%     \subfigure[$\theta_{\max}=\pi/3$]{
%         \includegraphics[width =0.3\textwidth]{test_L2_local_sample_vary_N2400_pi3.png}}
%     \subfigure[$\theta_{\max}=\pi/4$]{
%         \includegraphics[width =0.3\textwidth]{test_L2_local_sample_vary_N2400_pi4.png}}
%     \subfigure[$\theta_{\max}=\pi/8$]{
%         \includegraphics[width =0.3\textwidth]{test_L2_local_sample_vary_N2400_pi8.png}}
%     \caption{Average $l_2$-error against the Dirichlet distribution parameter $\alpha$ for the proposed approach DIR (solid-star), SLS (dash-circle), PLS (dash-triangle), DRD (dot-dash-square), and CIR (dot-dash-diamond) under different levels of angular similarity: $\theta_{\max}=\pi/3$ (a), $\theta_{\max}=\pi/4$ (b), and $\theta_{\max}=\pi/8$ (c). The simulations fix $N=2400$, $p=20$ and $m=30$.}
%     \label{Figure: Effect of local sample size in the Dirichlet}
% \end{figure}

\noindent{\bf Intensity of Noise.} 
To evaluate the impact of noise variance on estimation performance, we fix $N=2400$, $p=20$, and $m=30$. The local measurement size follows a power-law distribution, as illustrated in Figure \ref{Figure:DataDistribution} .  
The noise intensity and sign-flip probability $(\sigma_j,q_j)$ are configured to take values from $\{(0.1,0.75),(0.2+0.4\times(k-1),0.125)\}$ with equal probability $1/2$, where $k=1,\ldots,6$ controls the intensity of noise. This setup creates progressively noisier conditions as k increases. 

Figure \ref{Figure: Heterogeneity of Noise} shows the average $l_2$-error across nodes for $\theta_{\max}\in\{\pi/3,\pi/4,\pi/8\}$ in Appendix \ref{app:suppResult}. 
Our proposed approach demonstrates consistent performance stability as $k$ increases, maintaining accuracy levels comparable to the centralized method while outperforming other approaches. This robustness to increasing noise heterogeneity is further supported by additional results presented in Table \ref{Table:HeterNoise} and Figure \ref{Figure:HeterNoiseCos} in Appendix \ref{app:suppResult}.

\noindent{\bf Probability of Sign-flips.}
Finally, we investigate the effect of the probability of sign-flips by fixing $N=2400$, $p=20$, and $m=30$. The local measurement sizes follow a power-law distribution, as Figure \ref{Figure:DataDistribution} in Appendix \ref{app:suppResult} shows. Here $(\sigma_j,q_j)$ takes values from $\{(0.1,0.75),(0.2,0.075+0.025\times(k-1))\}$ with equal probability, where $k$ controls the degree of sign-flipping probability variation. Figure \ref{Figure: Heterogeneity of sign-flips} in Appendix \ref{app:suppResult} plots the average $l_2$-error across nodes. Our proposed method consistently matches the centralized method CIR and outperforms PLS, SLS and DRD. More detailed results are given in Table \ref{Table:HeterSign} and Figure \ref{Figure:HeterSignCos} in Appendix \ref{app:suppResult}.

\section{Application to reconstruct EEG signals using SEED dataset}\label{sec:application}
%\subsection{Dataset overview and preprocessing}
In this section, we validate our proposed method using the SJTU Emotion EEG Dataset (SEED) \citep{zheng2015investigating}, a widely recognized and comprehensive dataset for emotion recognition research, developed and maintained by the Brain-Computer Interface and Machine Intelligence Laboratory at Shanghai Jiao Tong University. The dataset is publicly accessible at \url{https://bcmi.sjtu.edu.cn/home/seed/}.

The SEED dataset comprises EEG signals collected from fifteen subjects while they watch carefully selected movie clips designed to evoke specific emotional states, including positive, negative, and neutral emotions. Each subject participated in 15 trials, with neural activity recorded using a 62-channel EEG electrode system. Detailed electrode placement information is provided in the literature \citep{zheng2015investigating}.

\begin{figure}[!htbp]
    % \flushleft %左对齐
    \centering
    \subfigure[]{
        \includegraphics[width =0.4\textwidth]{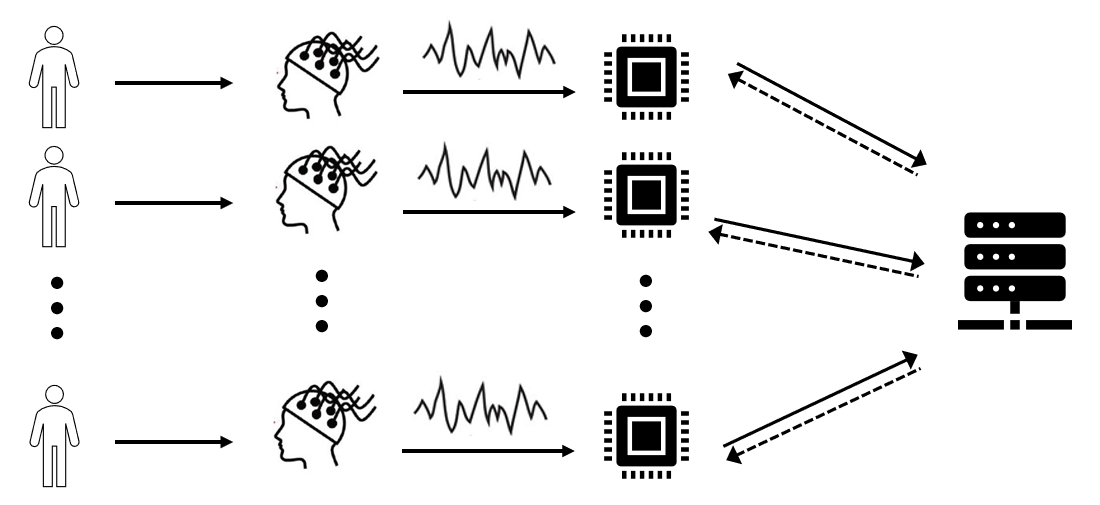}}
    \quad
    \subfigure[]{
        \includegraphics[width =0.4\textwidth]{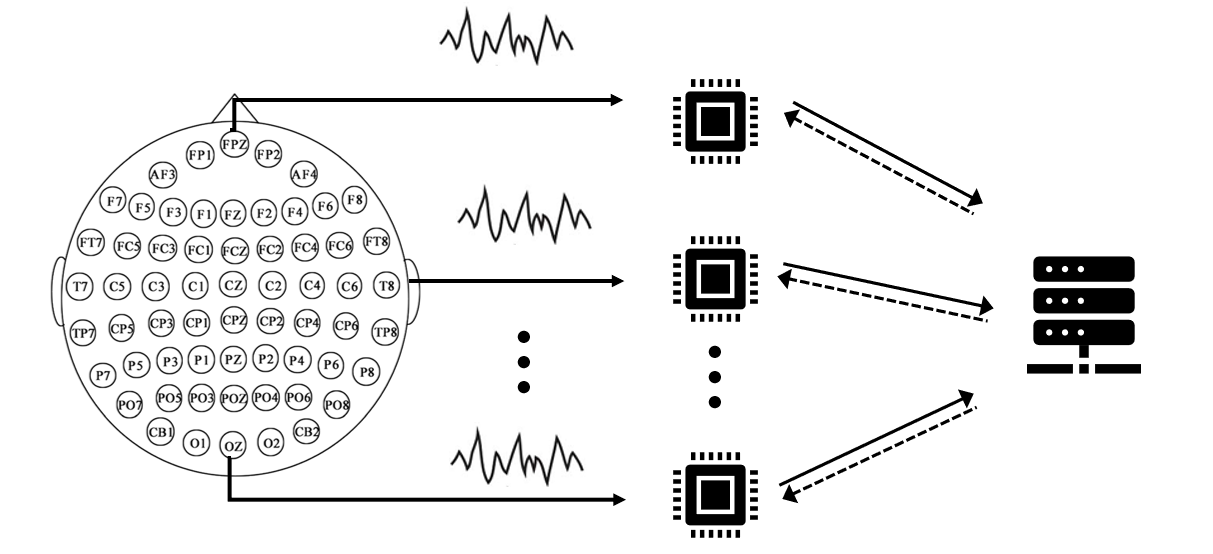}}
    \caption{The distributed setting for the application to SEED Datasets.
    }
    \label{Figure:EmpiricalStudySketch}
\end{figure}

\subsection{1-bit measurement generation}
The raw EEG signals were sampled at 1 kHz during approximately 4-minute film stimuli, yielding around 240,000 sample points per trial. We apply a bandpass filter (0.5–50 Hz) to remove noise and artifacts, followed by downsampling to 200 Hz, suitable for analyzing signals below 50 Hz. This results in EEG data sampled at 200 Hz with a frequency range of 0.5–50 Hz, temporally aligned with the video clips. To standardize signal lengths across trials, we extract 0.2-second and 1-second segments per trial, corresponding to signal dimensions of $p=40$ and $p=200$, respectively.

The preprocessed EEG signals are transformed into 1-bit measurements for distributed reconstruction. For each $p$-dimensional signal, with $p$ equal to $40$ or $200$, 
compression is performed using a node-specific measurement matrix $\X_j\in \mR^{d\times n_j}$. Each column vector of $\X_j$ is randomly drawn from a multivariate normal distribution $\mathcal{N}({\bf 0,\Sigma})$, with $\bSig=0.3^{|k-l|}$. To account for wireless channel imperfections, we introduce noise and sign-flip effects by setting the noise variance and flip probability parameters \( (\sigma_j, q_j) \) to \( \{(0.1, 0.75), (0.95, 0.025)\} \) with equal probability. The final 1-bit measurements are generated based on \eqref{eq:model}. The complete process is illustrated in Figure \ref{Figure:EmpiricalStudyProcess} in Appendix \ref{app:suppResult}.

\subsection{Experiment setup}
The SEED dataset exhibits two key heterogeneous patterns with latent similarities: (1) inter-subject similarity, where signals from the same electrode channel across different subjects show coherence under identical stimuli, and (2) intra-subject similarity, where signals from different electrode channels within the same subject exhibit coherence during identical stimuli. To systematically investigate these patterns and their implications for distributed signal processing, we designed two complementary experiments.

\begin{figure}[!htbp]
    % \flushleft %左对齐
    \centering
    \subfigure[Channel FP1]{
        \includegraphics[width =0.3\textwidth]{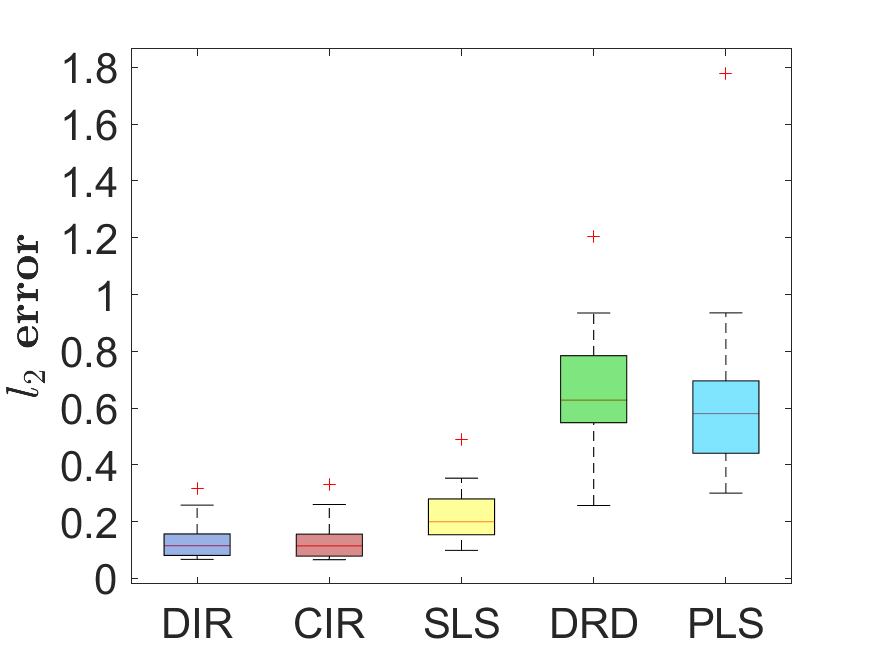}}
    \subfigure[Channel AF3]{
        \includegraphics[width =0.3\textwidth]{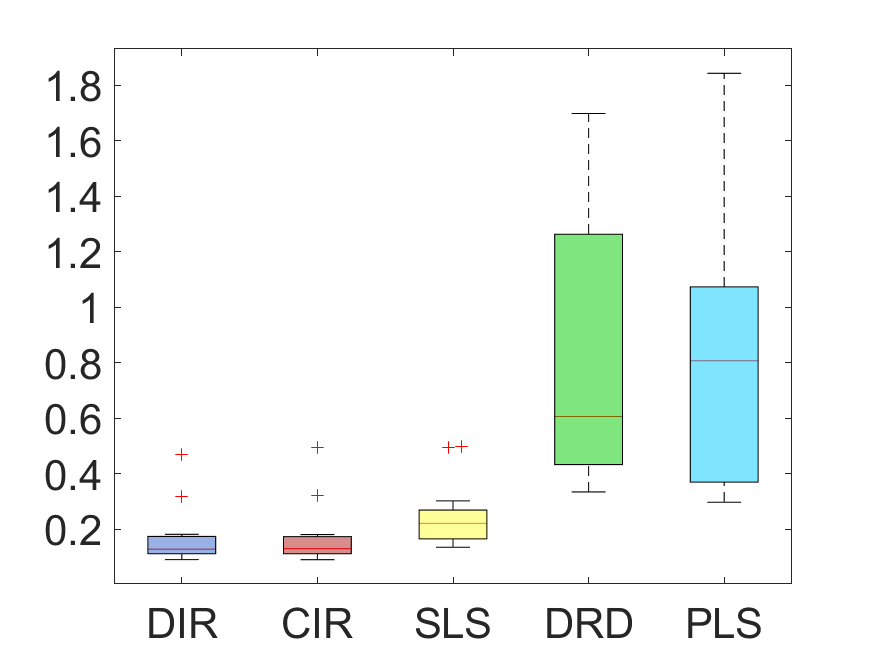}}
    \subfigure[Channel FZ]{
        \includegraphics[width =0.3\textwidth]{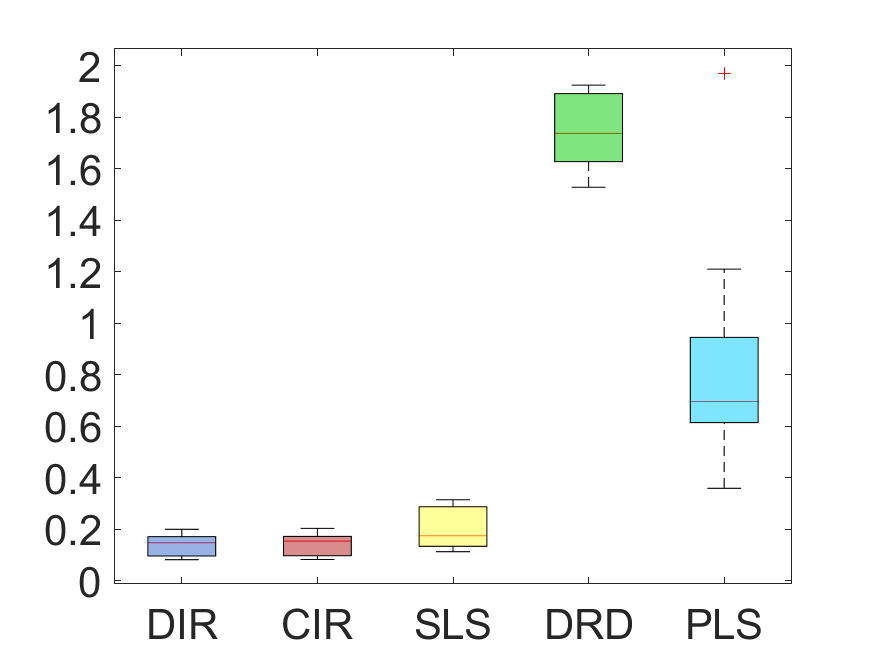}}
    \caption{Distribution of $l_2$-error across nodes for the proposed approach DIR (purple), CIR  (red), SLS (yellow), DRD (green), and  PLS(blue) 
    under different channels FP1 (a), AF3 (b) and FZ (c), with $m=15$, $p=40$ and $N=4500$. The samples follow a Dirichlet distribution with $\alpha = 0.7$. The real signal is a $0.2s$ segmented EEG data recorded by the above each channel from $15$ subjects under the same film clips.
    }
    \label{Figure: empirical study one L2}
\end{figure}

\noindent{\bf Experiment 1 (Multi-Subject Single-Channel Analysis).} 
We analyze EEG signals recorded from the same electrode channel across 15 subjects exposed to identical video clips, focusing on three channels: FP1, AF3, and FZ. The experimental setup, illustrated in Figure \ref{Figure:EmpiricalStudySketch} (a), includes a signal segmentation of 0.2s windows, 
a network of $m=15$ nodes, and a signal dimension of $p=40$. We consider two scenarios for the number of 1-bit measurements per node: (1) a Dirichlet distribution with $\alpha=0.7$, totaling $N=4500$ measurements, and (2) a uniform distribution with $n_j=300$ measurements per node.   

\noindent{\bf Experiment 2 (Single-Subject Multi-Channel Analysis).} 
We examine EEG signals from all 62 channels within a single subject across three emotional conditions (negative, neutral, positive), as depicted in Figure \ref{Figure:EmpiricalStudySketch} (b). The experimental setup includes signal segmentation of 1s windows, a network of $m=62$ nodes, and a signal dimension of $p=200$. Two scenarios are evaluated for the number of 1-bit measurements per node: (1) a Dirichlet distribution with $\alpha=0.7$, resulting in a total of $N=93,000$ measurements, and (2) a uniform distribution with $n_j=1500$ measurements per node. 

\begin{figure}[!htbp]
    % \flushleft %左对齐
    \centering
    \subfigure[Negative]{
        \includegraphics[width =0.3\textwidth]{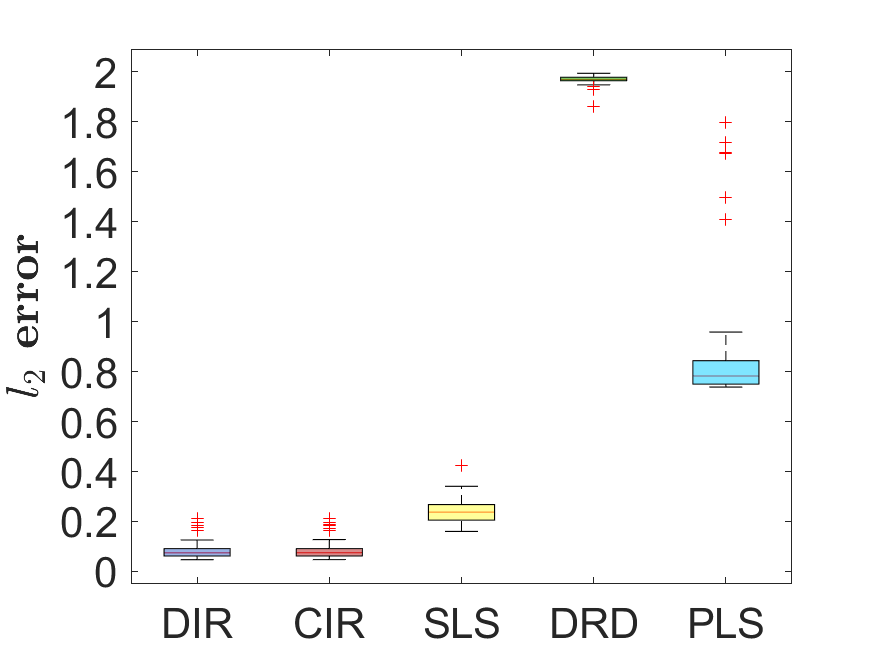}}
    \subfigure[Neutral]{
        \includegraphics[width =0.3\textwidth]{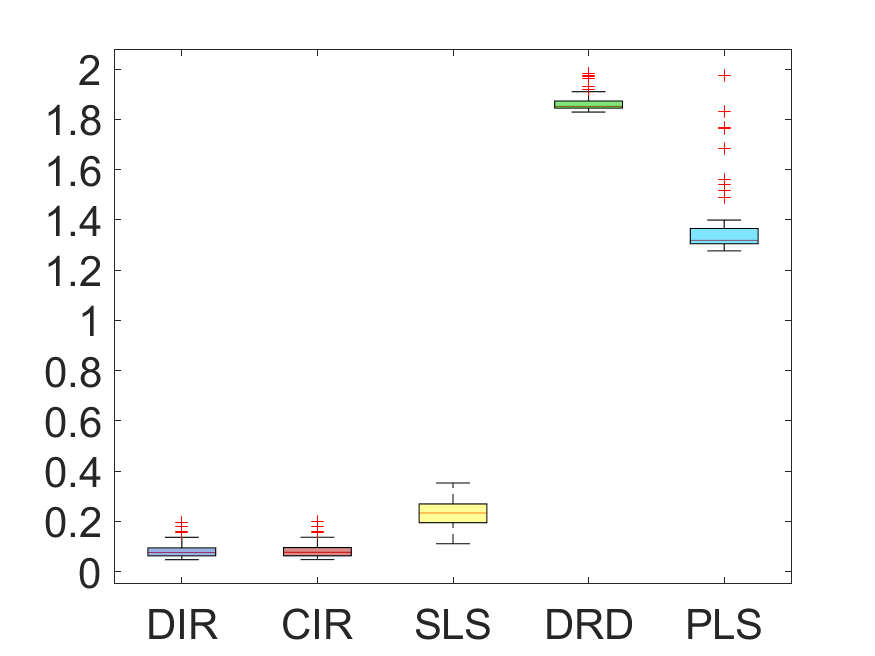}}
    \subfigure[Positive]{
        \includegraphics[width =0.3\textwidth]{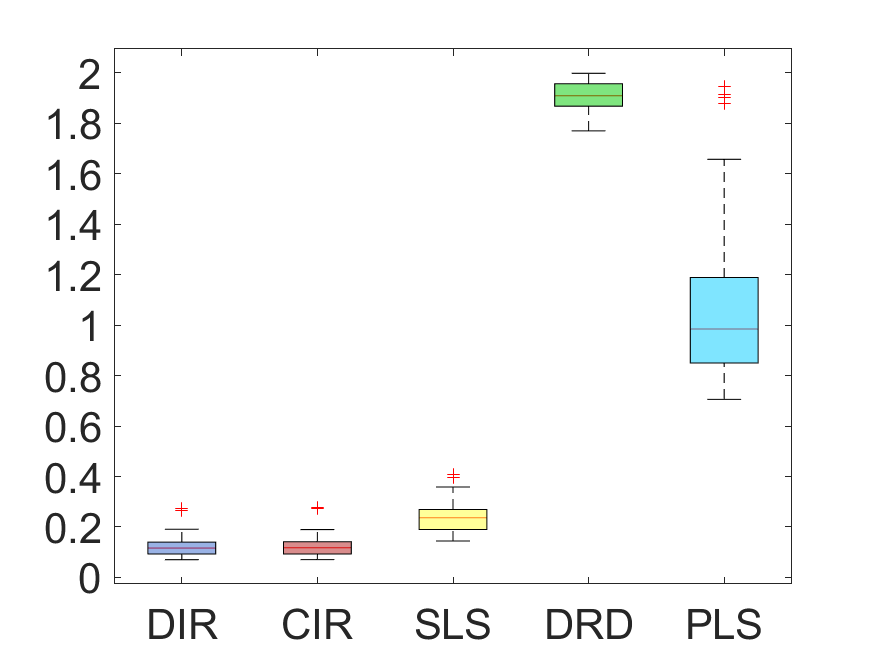}}
    \caption{Distribution of $l_2$-error across nodes for the proposed approach DIR (purple), CIR  (red), SLS (yellow), DRD (green), and  PLS(blue) 
    under different film clips negative (a), neutral (b), and positive (c), with $m=62$, $p=200$ and $N=93000$. The sample distribution is drawn from a Dirichlet distribution with $\alpha = 0.7$. The real signal is a $1s$ segmented EEG data from the same subject under each film clip, recorded by the 62-channel EEG electrode system.
    }
    \label{Figure: empirical study two L2}
\end{figure}

\subsection{Results}
The results, shown in Figures \ref{Figure: empirical study one L2} and \ref{Figure: empirical study two L2}, demonstrate the superior performance of our method in reconstructing EEG signals under both inter-subject and intra-subject similarity conditions. Compared to SLS, DRD, and PLS, our approach consistently improves the estimation performance for each node, maintaining consistency in both centralized and distributed settings. It achieves a higher reconstruction accuracy in both similarity patterns, signal dimensions, network sizes, and sample quantities. 
% A key advantage of our method is its ability to balance local sample information with inter-node dependencies while preserving communication efficiency. 
Unlike DRD, which requires sufficient local samples for PLS-comparable reconstruction, our method leverages node similarities to relax this constraint.
% Unlike DRD, which requires sufficient local samples for PLS-comparable reconstruction, our method leverages node similarities to relax this constraint, enhancing adaptability to diverse experimental conditions. 
Additional results with equal measurement quantities across nodes are provided in Appendix \ref{app:suppResult}, where Figures \ref{Figure:ESoneL2SameSample} and \ref{Figure:EStwoL2SameSample} show consistent trends.

\section{Conclusions}\label{sec:conclusion}
This study investigates the distributed reconstruction of heterogeneous signals with 1-bit CS measurement. The reconstruction challenge is formulated as a penalized optimization problem that balances accuracy and signal similarity. To address the nonconvexity, we introduce a novel invex relaxation technique, ensuring unique global optimality. This leads to a communication-efficient distributed algorithm that iteratively solves the problem by exchanging only gradient information.  To offer a clear and focused map to elucidate our method's mechanism for addressing heterogeneity, this work exclusively considers cases where the number of local samples exceeds the signal dimension, while 1-bit CS often assumes sparsity.

Theoretical analysis confirms the statistical consistency of the estimate under directional alignment of signals across nodes, leading to tighter reconstruction error bounds than local least squares methods. Moreover, global convergence theory demonstrates that the proposed algorithm progressively enhances the reconstruction accuracy with each iteration, ultimately attaining the minimax optimal rate. Numerical simulations validate the effectiveness and efficiency of the proposed method, providing empirical support for the theoretical findings. For future research, extending the work to handle high-dimensional sparse signals and incorporating network structures among nodes could further improve recovery performance. Furthermore, exploring robustness against noise and adversarial attacks, as well as the feasibility of real-time applications, would be valuable extensions of this work.
%-----------------------------------------------------------------------------------------------------
% \newpage

{
\centering
\normalem
\bibliographystyle{chicago}
\setlength{\bibsep}{0.5ex} 
\begin{spacing}{1.47}
\bibliography{ref}
\end{spacing}
}

\end{document}

\maketitle
%-----------------------------------------------------------------------------------------------------

\begin{appendix}

\section{The Lemma for Theorem \ref{Theorem:1} with Specific Factors} 
\renewcommand{\theequation}{A.1.\arabic{equation}}
\setcounter{equation}{0}

In this section, we provide the following lemmas corresponding to the results 
 in Theorem \ref{Theorem:1}  with specific factors.

Let $a_j=(2q_j-1)\{\pi(\sigma_j^2+1)/2\}^{-1/2}$, $a_{\min}=\min_{1\leq j\leq m}a_j$, and $a_{\max}=\max_{1\leq j\leq m}a_j$. Let $\kappa(\bSig) = \gamma_{\max}(\bSig)/\gamma_{\min}(\bSig)$ denote the condition number of the covariance matrix $\bSig$. 
To give the appropriate order of $\lambda$ for obtaining estimated performance improvement, we define
\begin{align}
    \lam_L =&  \left[  \alpha_1(m,r,\theta) +(3C_4a_{\min}^2a_{\max}^2pRN)^{-1}|\cos\theta|
    \right]^{-1}(1-r^{1/2}), \label{def:lamL} \\
    \lam_U = &  \alpha_1^{-1}(m,r,\theta)(1+r^{1/2}) ,\label{def:lamU}
\end{align}
where 
\begin{align}
    \alpha_1(m,r,\theta) =& {C_4\{r+(1-\cos\theta)\}}/{mrC_5}, \label{def:alpha1} 
\end{align}
and denote the following constants
\begin{align}
    \alpha_2(m,r,\theta,n_j) =& \alpha_1(m,r,\theta) + C_4/n_j,
    \label{def:alpha2}\\
    m_1 =&  a_{\min}C_{\max}^{-1/2}\kappa(\bSig)^{-1} /3  \label{def: m1},   \\
    m_2 =&  3p^{1/2}\kappa(\bSig)C_{\min}^{-1/2}a_{\max}, \label{def: m2}   \\
    C_3 =&  C_2^2/C_1, \label{def: C3}   \\
    C_4 =&  2C_2\kappa(\bSig)^3/a_{\min}^2, \label{def: C4} \\
    C_5 =&  4C_3^2\left\{\kappa^{1/2}(\bSig)+(6a_{\min})^{-1}\log^{1/2} p \right\}^2, \label{def: C5}
\end{align}
where $C_1$ and $C_2$  are constants independent from $p$ and $n_j$, for $j=1,...,m$.  $m_1$, $C_3$ and $C_4$ are some generic constants not depending on $p$ and $n_j$, for $j=1,...,m$.
According to the above conditions, one can give statistical consistency.

{\lem [Nonasymptotic error bound with specific constant results] 

For any node $j=1,\cdots,m$, under conditions (C1) - (C4), if $n_j>C_5(m\sqrt{r}\vee d)$  and $|\cos\theta|\ge ({C_5p}/{n_j})^{1/2} $
 \begin{equation}
        \begin{split}
            \norm{\bbeta_j/a_j - \bbeta_j^*} \le& \left\{
        1 -  \lambda\alpha_1(m,r,\theta)\right\}\vee
        \left\{
        \lambda\alpha_2(m,r,\theta,n_j)- 1
        \right\} T_1(p/n_j)^{1/2}
        \\ &
        +  2\lambda \alpha_1(m,r,\theta) T_1 (p/N)^{1/2}
        +\lambda T_2d^{1/2}/n_j,
        \end{split}
        \label{lem: result 1}
\end{equation}
with probability at least  $1-O\left(m\exp(-C_1C_2^2p)+m/p^3\right)$, where $\alpha_1(m,r,\theta)$ and $\alpha_2(m,r,\theta)$ are defined  in \eqref{def:alpha1} and \eqref{def:alpha2}, and 
\begin{align}
    T_1 &=  (8/C_{\min})\left\{
                2C_2{(\kappa(\Sigma)C_{\max})^{1/2}} 
                + 3(|2q_j-1|C_1^{1/2})^{-1}(\sigma_j+1){\log^{1/2} p}
            \right\}, \label{eq:T1} \\
    T_2 &= 6(|2q_j-1|C_{\min}m_1)^{-1}(\sigma_j+1)
            \left\{
                1- \left(C_2p^{1/2}\kappa(\Sigma)^3\right)^{-1}+{(2-2|\cos\theta|)^{1/2}}
            \right\}. \label{eq:T2}
\end{align}
Given $\lam_L \le \lambda \le \lam_U$, we have
\begin{equation}
    \norm{\bbeta_j/a_j - \bbeta_j^*} \le
    3r^{-1}T_1(p/N)^{1/2} + C_4^{-1}{C_5} T_2p^{1/2}/n_j.\label{lem: result 2}
\end{equation}
\label{lem:Specific_Error}
}

Lemma \ref{lem:Specific_Error} gives the specific factors of local quantization errors ($O\{(1-\lam)(p\log{p}/n_j)^{1/2}\}$), network-wide quantization errors ($O\{\lam(p\log{p}/N)^{1/2}\}$), and penalty function errors ($O(\lam p^{1/2}/n_j)$). Obviously, there is a trade-off between introducing inter-node information and network-wide quantization errors. The factor $O(\log p)$ of quantization errors can be eliminated by utilizing the inter-node information. Hence, as mentioned in the main text, with $\lam$ in the appropriate range, we can obtain the minimax optimal rate up to a constant factor.

%---------------------------------------------------------
\section{Supplementary experimental results}\label{app:suppResult}
\setcounter{lem}{0} 
\renewcommand{\thefigure}{C.\arabic{figure}}
\setcounter{figure}{0}
\renewcommand{\theequation}{C.\arabic{equation}}
\setcounter{equation}{0}
\renewcommand{\thetable}{C.\arabic{table}}
\setcounter{table}{0}

\subsection{Toy example for initialization robustness}
\begin{table}[!htbp]
\centering
\caption{The comparison of initialization robustness for original and invex formulation.}
\setlength{\tabcolsep}{4mm}{
\begin{tabular}{ccccccc}
\hline
\textbf{Initialization} & & \textbf{1} & \textbf{2} & \textbf{3} & \textbf{4} & \textbf{5} \\
\hline
\multirow{5}{*}{Invex Formulation} 
&Node 1 & 0.0789 & 0.0789 & 0.0790 & 0.0788 & 0.0793 \\
&Node 2 & 0.3066 & 0.3065 & 0.3065 & 0.3064 & 0.3069 \\
&Node 3 & 0.0614 & 0.0603 & 0.0601 & 0.0606 & 0.0601 \\
&Node 4 & 0.0040 & 0.0039 & 0.0040 & 0.0039 & 0.0040 \\
&Node 5 & 0.0257 & 0.0258 & 0.0257 & 0.0258 & 0.0256\\
\hline
\multirow{5}{*}{Original Formulation} 
&Node 1 & 0.1617    & 0.2528    & 0.6529    & 0.1869    & 0.5542 \\
&Node 2 & 1.3443 & 0.0999 & 0.7406 & 0.0714 & 0.8777 \\
&Node 3 & 0.7960 & 0.2092 & 0.6649 & 0.2228 & 1.2864 \\
&Node 4 & 1.9198 & 0.2477 & 0.4327 & 0.2977 & 0.1839 \\
&Node 5 & 0.3782 & 0.0941 & 1.6337 & 0.0691 & 1.0628 \\
\hline
\end{tabular}
}
\begin{tablenotes} 
\footnotesize
\item{$\bullet$ An example for 5 nodes with 2-dimensional signals and 10 measurements on each.}
\item{$\bullet$ We present the absolute cosine values of estimated and real signals on each node per initialization.}
\item{$\bullet$ For invex Formulation, we conduct the DIR algorithm. And for Original Formulation, we conduct gradient descent directly.}
\end{tablenotes} 
\label{tab:initialization}
\end{table}

\subsection{Performance Results for Algorithm}

\begin{figure}[!htbp]
    \centering
        \includegraphics[width =0.5\textwidth]{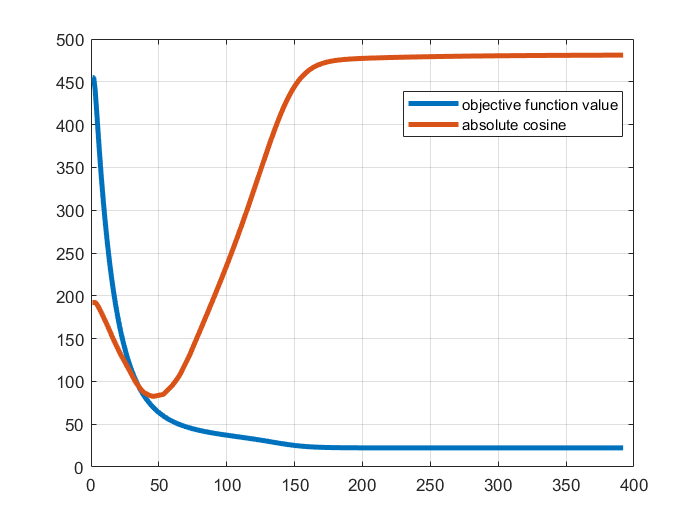}
    \caption{ The convergence of algorithm with $m=30$, $p=20$, $N=2400$ and $\theta_{\max} = \pi/8$}
    \label{Figure:converge}
\end{figure}

\noindent{\bf The convergence of Algorithm \ref{Algorithm:1}.}
Figure \ref{Figure:converge} illustrates the evolution of the objective function value and absolute cosine between the estimated and true signals during communication rounds for a random replication under $N=2400$, $m=30$, $p=20$, and $\theta_{\max}=\pi/8$, where we let the cosine similarity between the signals at any two nodes be greater than $\cos\theta_{\max}$. The objective function value shows a decreasing trend as the algorithm progresses, indicating successful minimization.

\begin{figure}[!htbp]
    % \flushleft %左对齐
    \centering
    \subfigure[$\theta_{\max} = \pi/3$]{
        \includegraphics[width =0.45\textwidth]{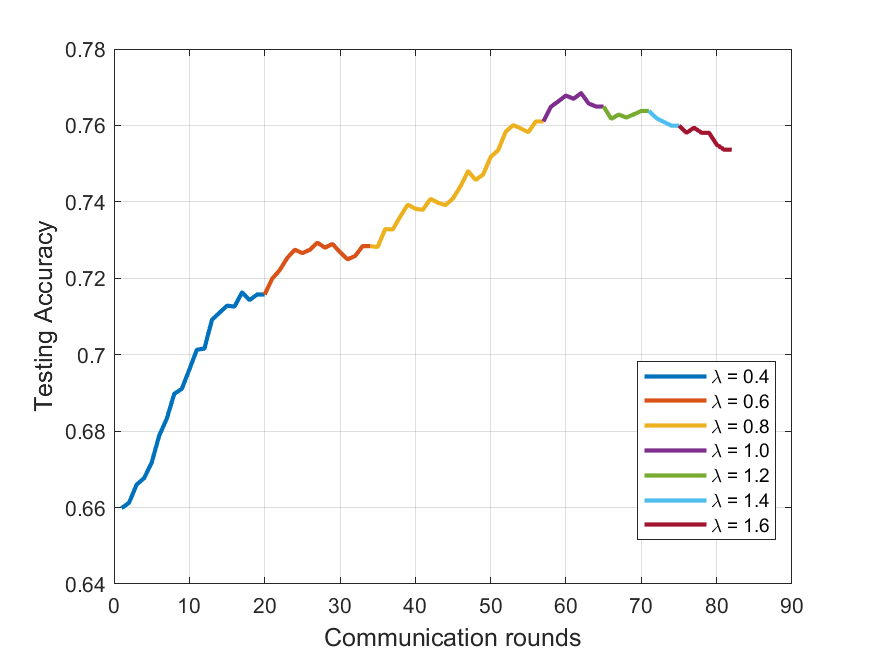}}
        \subfigure[$\theta_{\max} = \pi/4$]{
        \includegraphics[width =0.45\textwidth]{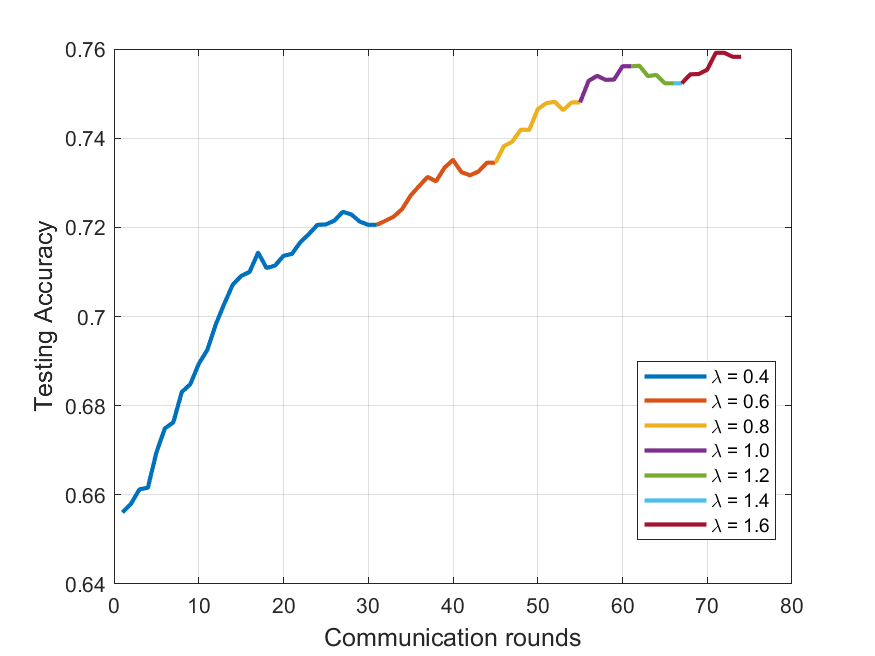}}
    \caption{The warming start strategy for regularization parameter selection on dataset, where $N=2400$, $m=30$, $p=20$, and local measurement size is given as Figure \ref{Figure:DataDistribution} which follows power law distribution.} 
    \label{Figure:Warming-up}
\end{figure}

\noindent{\bf The Warm-Up tuning strategy.} 
Figure \ref{Figure:Warming-up} shows how the validation accuracy varies with the regularization parameter $\lam$.

\noindent{\bf The running time.}
The running time of the warm-up strategy is given in \ref{Table:TuningTime}, compared to the separate tuning.
\begin{table}[!htbp]
    \centering
        \caption{Selected $\lambda$ by two strategies and running time on the same dataset used in Figure \ref{Figure:Warming-up}}
    \begin{tabular}{c ccccc }
    \hline
        &  Round & Strategy & running time (second)\\
    \hline
    \multirow{2}{*}{$\pi/3$} & \multirow{2}{*}{60} & warmup & 5.4034\\
    & & separate &  49.0019\\
    \multirow{2}{*}{$\pi/4$} & \multirow{2}{*}{65} & warmup & 5.9449\\
    & & separate &  53.3344\\
    \hline
    \end{tabular}
    \label{Table:TuningTime}
\end{table}

\subsection{More figures and tables}
\subsubsection{ Numerical Simulation}
\begin{figure}[!htbp]
    % \flushleft %左对齐
    \centering
        \includegraphics[width =0.5\textwidth]{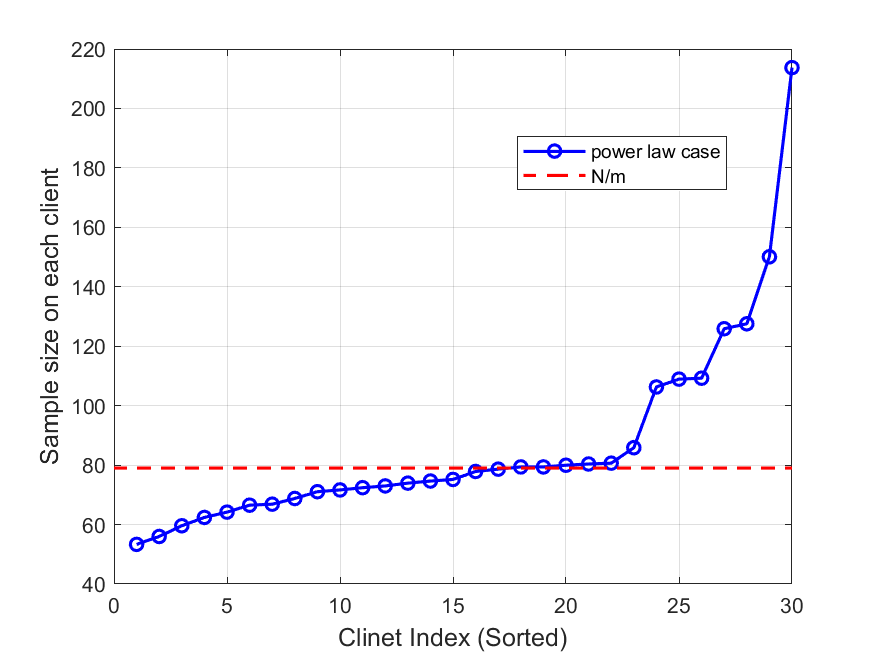}
    \caption{The data distribution following power law.}
    \label{Figure:DataDistribution}
\end{figure}

\noindent{\bf The data distribution with a power law.} 
Figure \ref{Figure:DataDistribution} represents the sample size distribution that we used for some experiments.

\begin{figure}[!htbp]
    % \flushleft %左对齐
    \centering
    \subfigure[$\theta_{\max}=\pi/3$]{
        \includegraphics[width =0.3\textwidth]{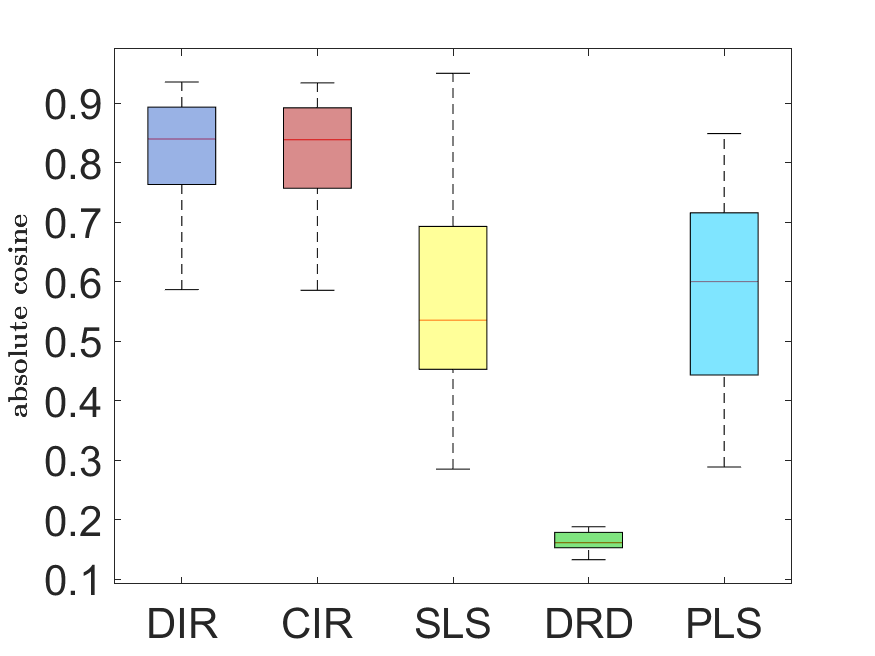}}
    \subfigure[$\theta_{\max}=\pi/4$]{
        \includegraphics[width =0.3\textwidth]{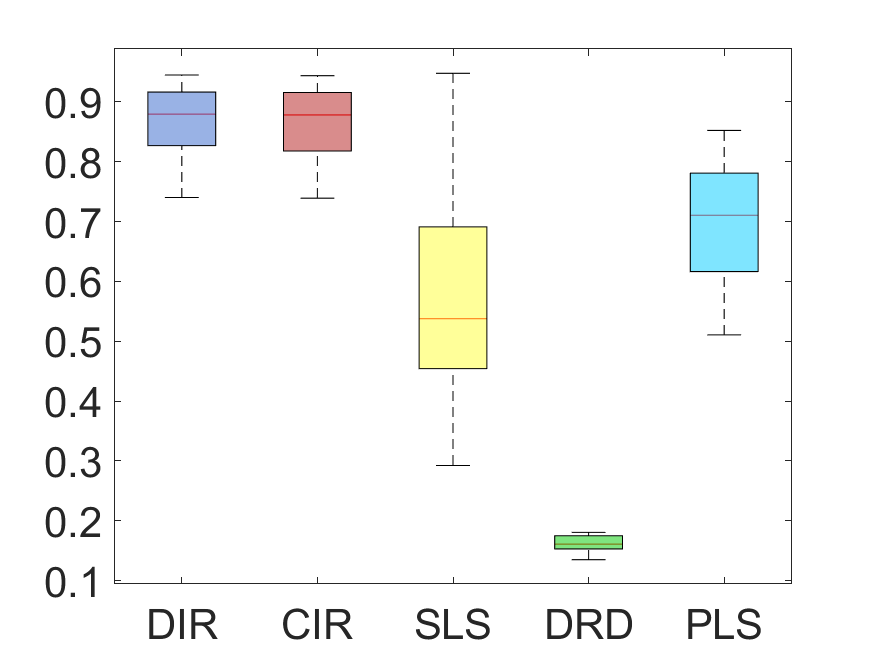}}
    \subfigure[$\theta_{\max}=\pi/8$]{
        \includegraphics[width =0.3\textwidth]{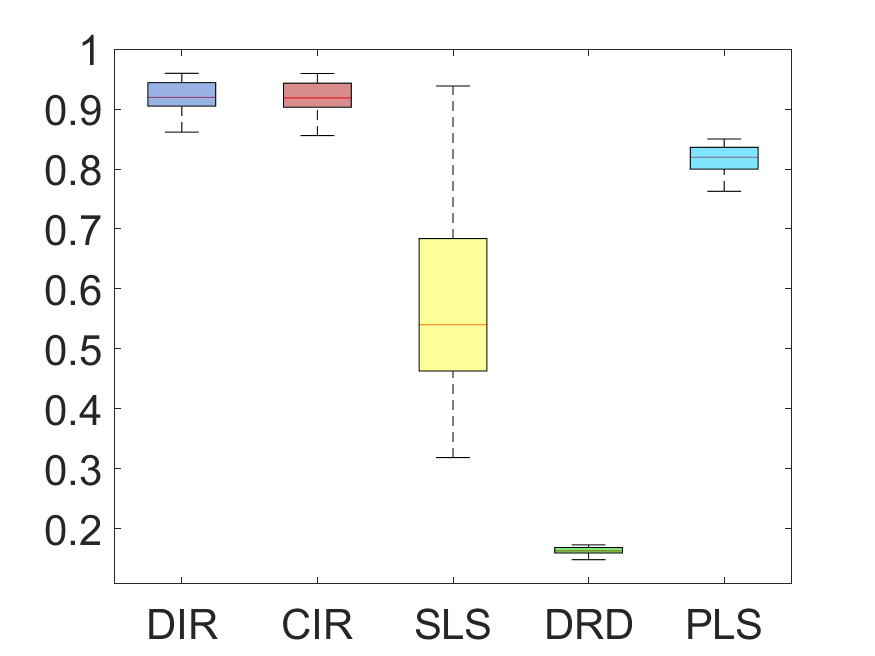}}
    \caption{Distribution of absolute cosine across nodes for the proposed approach (purple), SLS (red), PLS (yellow), DRD (green), and CIR (blue) under 
    different levels of angular similarity: $\theta_{\max}=\pi/3$ (a), $\theta_{\max}=\pi/4$ (b), and $\theta_{\max}=\pi/8$ (c). The simulations fix $N=2400$, $p=20$ and $m=30$.}
    \label{Figure:DistributionAcrossNodes2}
\end{figure}

\noindent{\bf The distribution and average of absolute cosine values across the nodes.} Figure \ref{Figure:DistributionAcrossNodes2} depicts the distribution of absolute cosine values between estimated parameters and true parameters across all nodes for datasets with $N=2400$, $m=30$, and $\theta_{\max}\in\{\pi/3,\pi/4,\pi/8\}$.

\begin{figure}[!htbp]
    % \flushleft %左对齐
    \centering
    \subfigure[$\theta_{\max}=\pi/3$]{
        \includegraphics[width =0.3\textwidth]{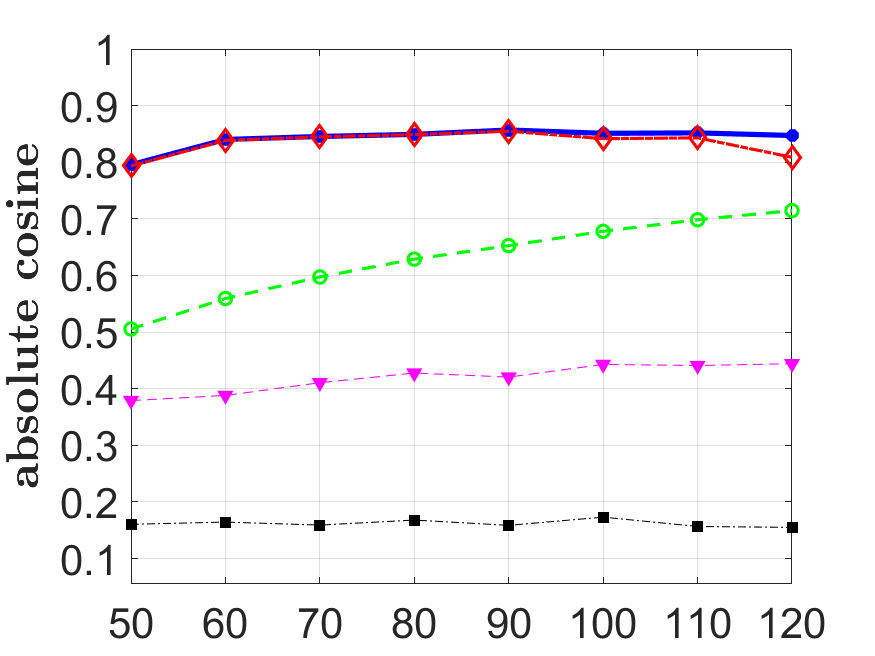}}
    \subfigure[$\theta_{\max}=\pi/4$]{
        \includegraphics[width =0.3\textwidth]{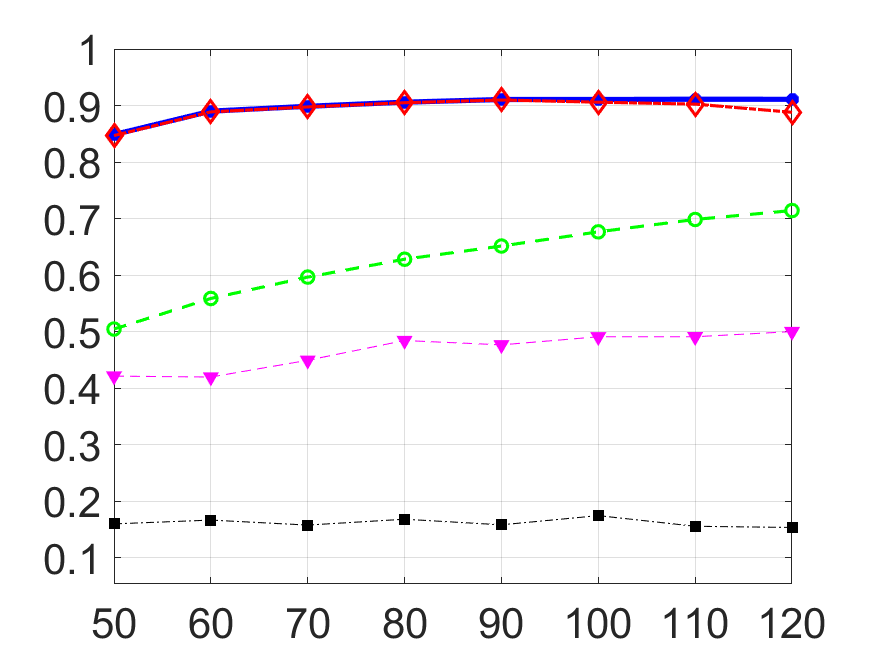}}
    \subfigure[$\theta_{\max}=\pi/8$]{
        \includegraphics[width =0.3\textwidth]{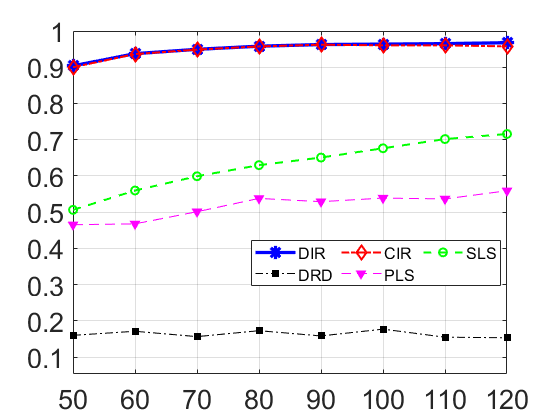}}
    \caption{Average absolute cosine against the number of local measurements for the proposed approach (solid-star), SLS (dash-circle), PLS (dash-triangle), DRD (dot-dash-square), and CIR (dot-dash-diamond) under 
    different levels of angular similarity: $\theta_{\max}=\pi/3$ (a), $\theta_{\max}=\pi/4$ (b), and $\theta_{\max}=\pi/8$ (c). The simulations fix $p=20$ and $m=30$.}
    \label{Figure: Effect of local sample size cos}
\end{figure}

\noindent{\bf The effect of the number of local measurements.} 
To test the effect of the number of local measurements, we fix the number of nodes  $m=30$, and vary the number of local measurements $n_j\equiv n\in\{50,60, \ldots, 120\}$. Figure \ref{Figure: Effect of local sample size cos} shows the   average absolute cosine result.

\begin{figure}[!htbp]
    % \flushleft %左对齐
    \centering
    \subfigure[$\theta_{\max}=\pi/3$]{
        \includegraphics[width =0.3\textwidth]{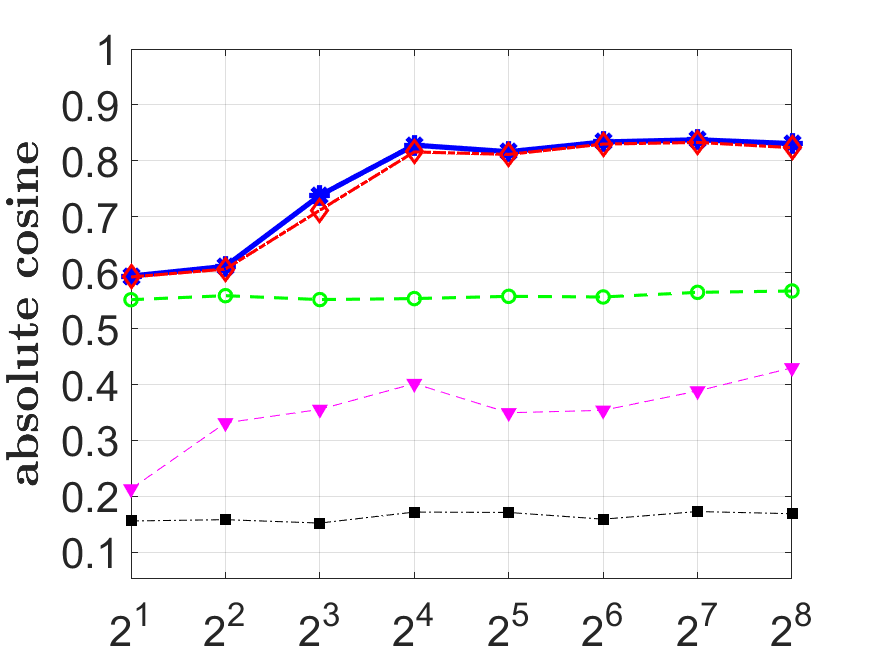}}
    \subfigure[$\theta_{\max}=\pi/4$]{
        \includegraphics[width =0.3\textwidth]{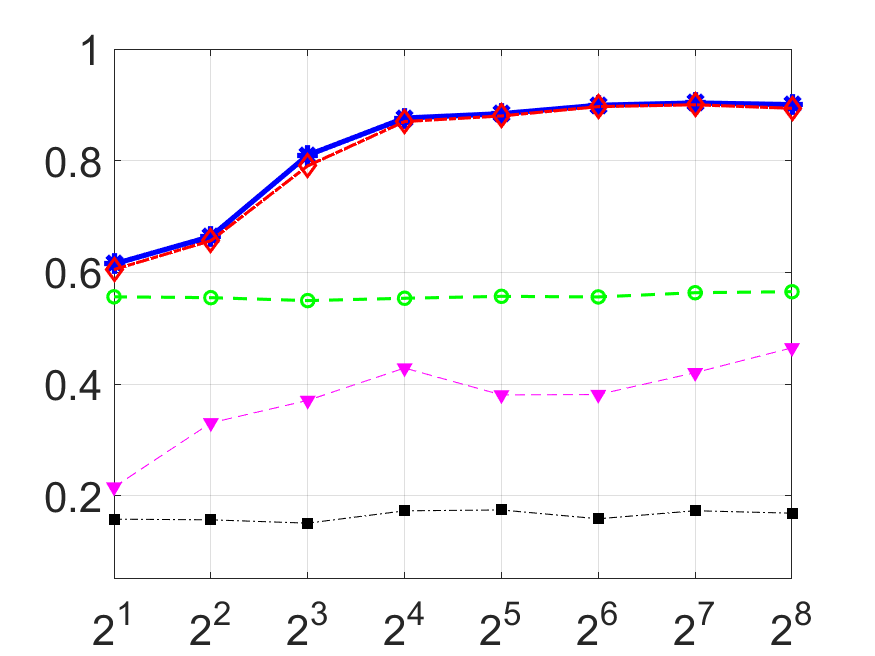}}
    \subfigure[$\theta_{\max}=\pi/8$]{
        \includegraphics[width =0.3\textwidth]{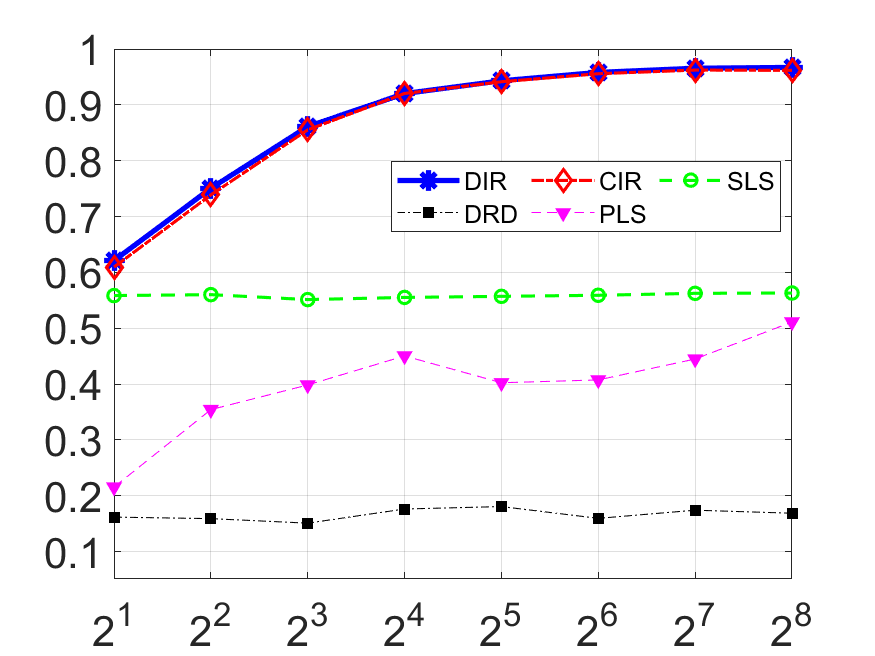}}
    \caption{Average absolute cosine against the base-2 logarithm number of nodes for the proposed approach (solid-star), SLS (dash-circle), PLS (dash-triangle), DRD (dot-dash-square), and CIR (dot-dash-diamond) under 
    different levels of angular similarity: $\theta_{\max}=\pi/3$ (a), $\theta_{\max}=\pi/4$ (b), and $\theta_{\max}=\pi/8$ (c). The simulations fix $p=20$ and $n=60$.}
    \label{Figure: m vary cos}
\end{figure}

\noindent{\bf The effect of varying the number of nodes $m$.} 
We  fix the local measurement size $n_j\equiv n=60$ and $m$ varies in $\{2^1,2^2,\ldots,2^{10}\}$. Figure \ref{Figure: m vary cos}  shows the   average absolute cosine result.

\begin{figure}[!htbp]
    % \flushleft %左对齐
    \centering
    \subfigure[$\theta_{\max}=\pi/3$]{
        \includegraphics[width =0.3\textwidth]{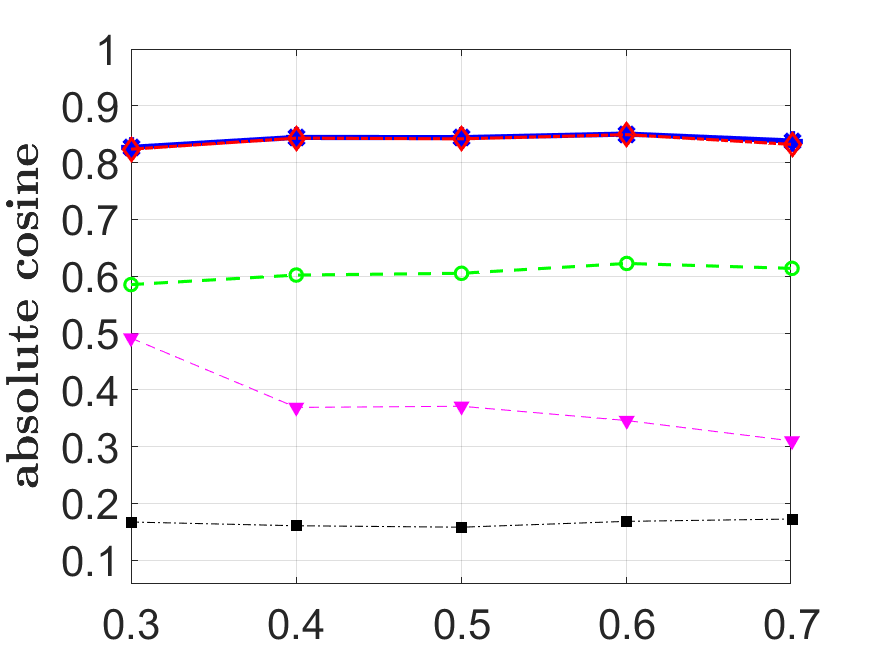}}
    \subfigure[$\theta_{\max}=\pi/4$]{
        \includegraphics[width =0.3\textwidth]{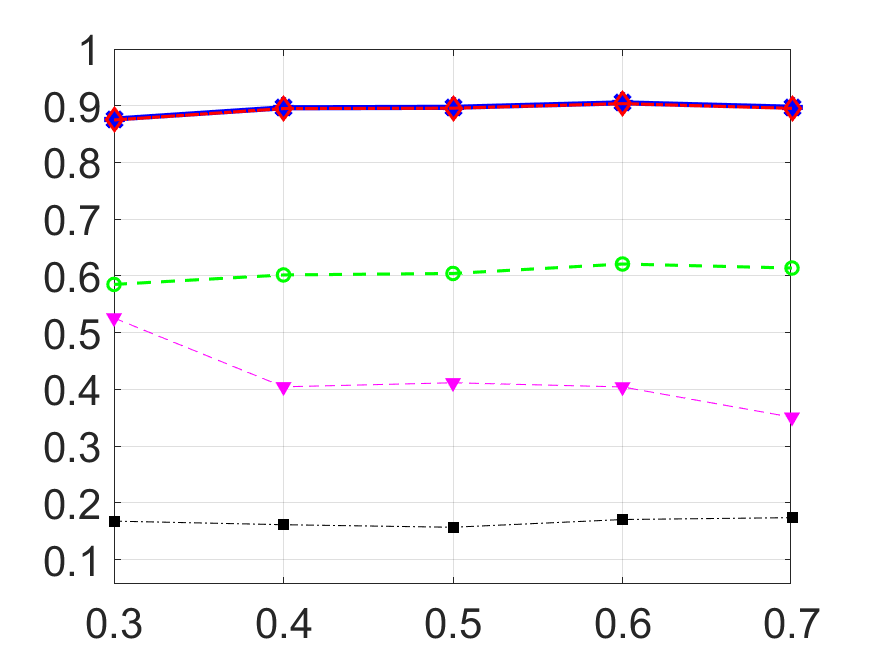}}
    \subfigure[$\theta_{\max}=\pi/8$]{
        \includegraphics[width =0.3\textwidth]{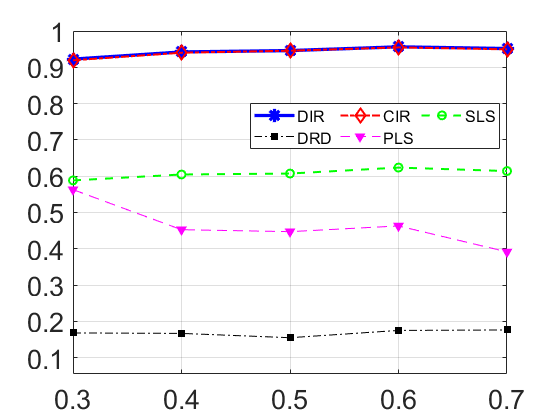}}
    \caption{Average absolute cosine against the parameter $\alpha$ of Dirichlet distribution for the proposed approach (solid-star), SLS (dash-circle), PLS (dash-triangle), DRD (dot-dash-square), and CIR (dot-dash-diamond) under 
    different levels of angular similarity: $\theta_{\max}=\pi/3$ (a), $\theta_{\max}=\pi/4$ (b), and $\theta_{\max}=\pi/8$ (c). The simulations fix $N=2400$, $p=20$ and $m=30$.}
    \label{Figure: Effect of local sample size in the Dirichlet cos}
\end{figure}

\noindent{\bf The effect of varying the number of nodes $m$.} 
Figure \ref{Figure: Effect of local sample size in the Dirichlet cos} shows the absolute cosines in the scenarios where the number of measurements follows
a Dirichlet($\alpha$) distribution with $\alpha\in {0.3, 0.4, 0.5, 0.6, 0.7}$.

\begin{figure}[!htbp]
    % \flushleft %左对齐
    \centering
    \subfigure[$\theta_{\max}=\pi/3$]{
        \includegraphics[width =0.3\textwidth]{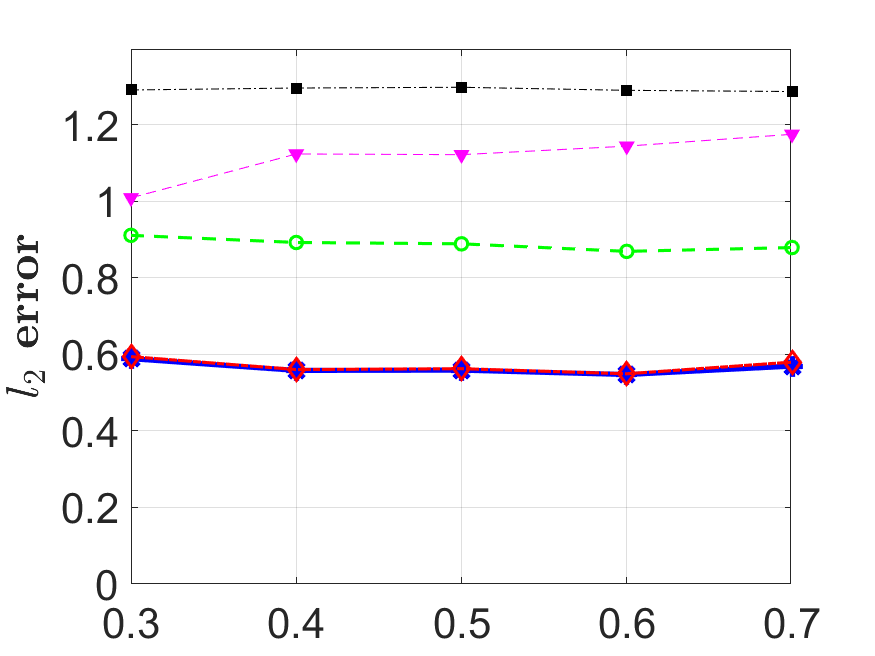}}
    \subfigure[$\theta_{\max}=\pi/4$]{
        \includegraphics[width =0.3\textwidth]{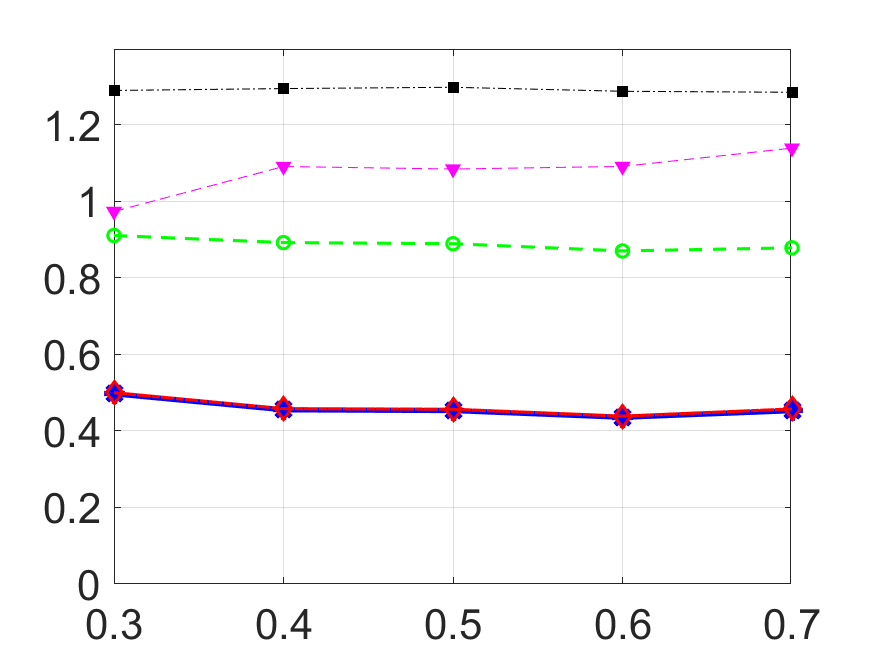}}
    \subfigure[$\theta_{\max}=\pi/8$]{
        \includegraphics[width =0.3\textwidth]{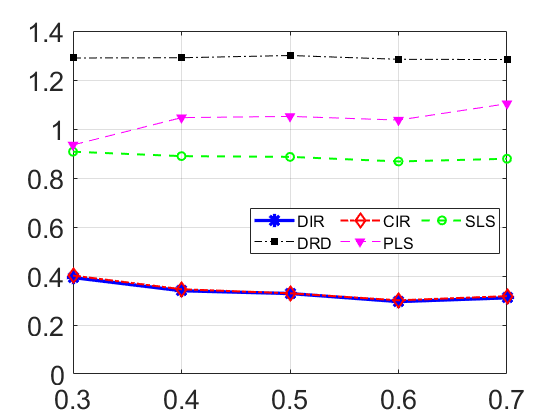}}
    \caption{Average $l_2$-error against the Dirichlet distribution parameter $\alpha$ for the proposed approach DIR (solid-star), SLS (dash-circle), PLS (dash-triangle), DRD (dot-dash-square), and CIR (dot-dash-diamond) under different levels of angular similarity: $\theta_{\max}=\pi/3$ (a), $\theta_{\max}=\pi/4$ (b), and $\theta_{\max}=\pi/8$ (c). The simulations fix $N=2400$, $p=20$ and $m=30$.}
    \label{Figure: Effect of local sample size in the Dirichlet}
\end{figure}

\begin{figure}[!htbp]
    % \flushleft %左对齐
    \centering
    \subfigure[$\theta_{\max}=\pi/3$]{
        \includegraphics[width =0.3\textwidth]{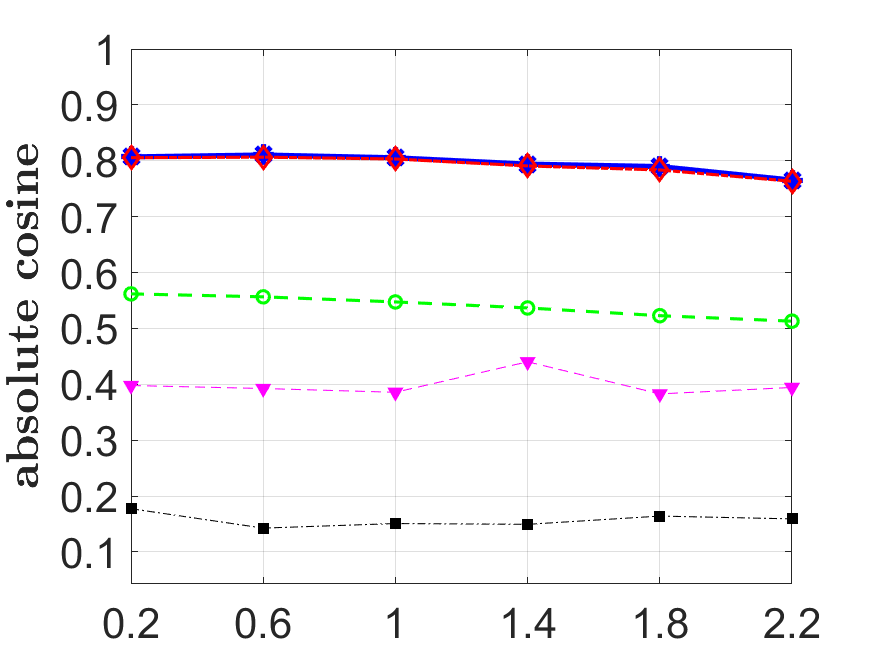}}
    \subfigure[$\theta_{\max}=\pi/4$]{
        \includegraphics[width =0.3\textwidth]{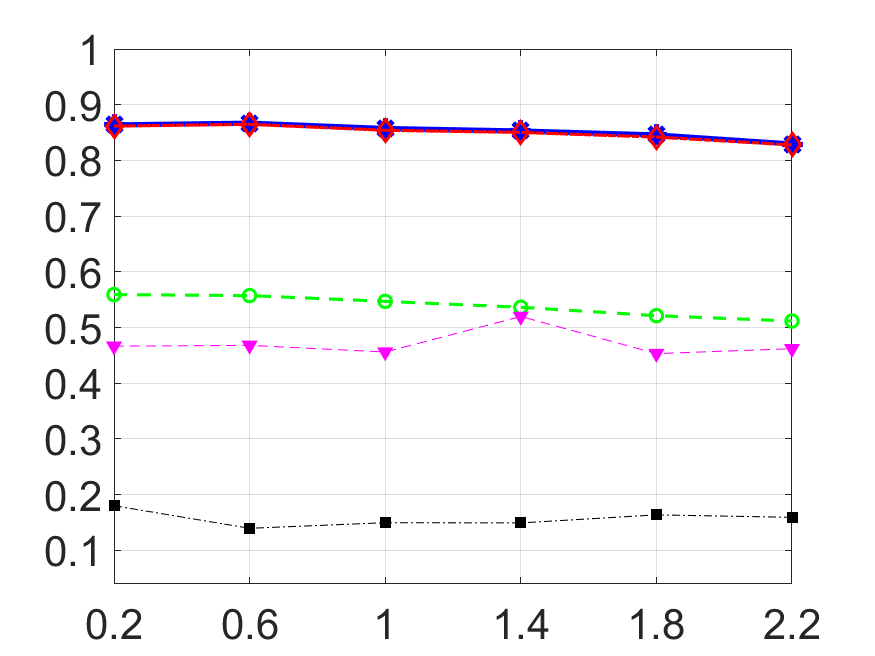}}
    \subfigure[$\theta_{\max}=\pi/8$]{
        \includegraphics[width =0.3\textwidth]{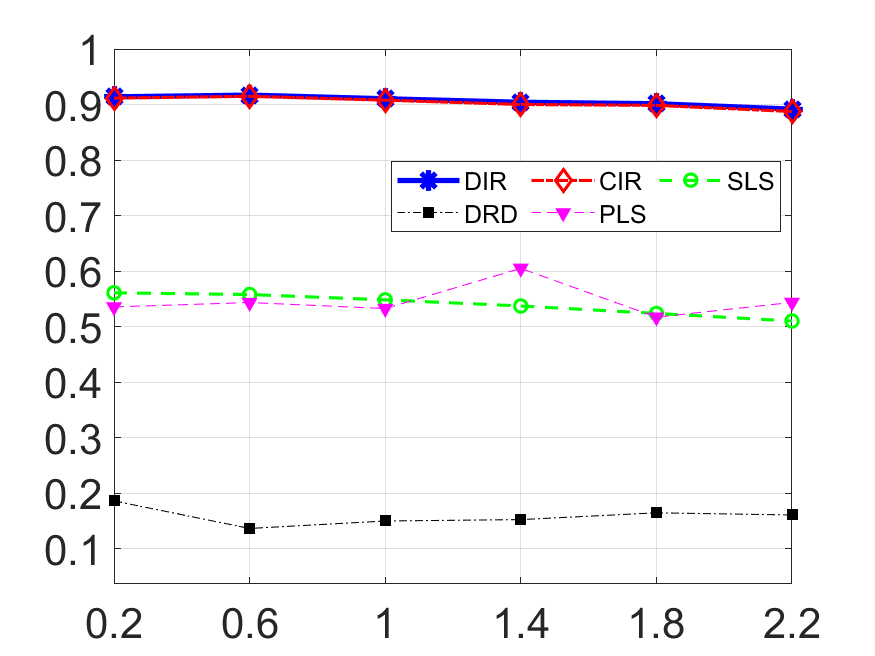}}
    \caption{Average absolute cosine against the intensity parameter $k$ of noise for the proposed approach (solid-star), SLS (dash-circle), PLS (dash-triangle), DRD (dot-dash-square), and CIR (dot-dash-diamond) under 
    different levels of angular similarity: $\theta_{\max}=\pi/3$ (a), $\theta_{\max}=\pi/4$ (b), and $\theta_{\max}=\pi/8$ (c). The simulations fix $N=2400$, $p=20$ and $m=30$, and let $(\sigma_j,q_j)$ take the values $\{(0.1,0.75),(0.2+0.4\times(k-1),0.125)\}$ with equal probability $1/2$, where $k=1,\ldots,6$ controls the intensity of noise.}
    \label{Figure:HeterNoiseCos}
\end{figure}

\noindent{\bf The specific result for noise intensity.} In Table \ref{Table:HeterNoise}, we present the $1/4$, $1/2$ and $3/4$ quantile of the absolute cosine of nodes for SLS and our method (Invex), with $\theta_{\max}= \pi/3$ and noise intensity varying. Improved Ratio represents the ratio of nodes that gain estimate improvement with our method compared to SLS.
Figure \ref{Figure:HeterNoiseCos} shows the absolute cosine values in these scenarios.

\begin{figure}[!htbp]
    % \flushleft %左对齐
    \centering
    \subfigure[$\theta_{\max}=\pi/3$]{
        \includegraphics[width =0.3\textwidth]{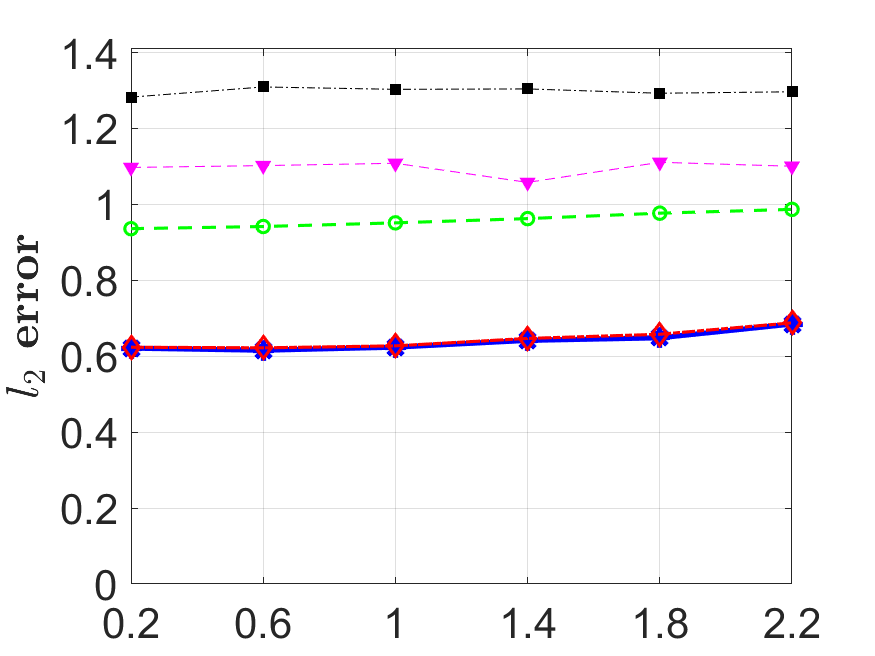}}
    \subfigure[$\theta_{\max}=\pi/4$]{
        \includegraphics[width =0.3\textwidth]{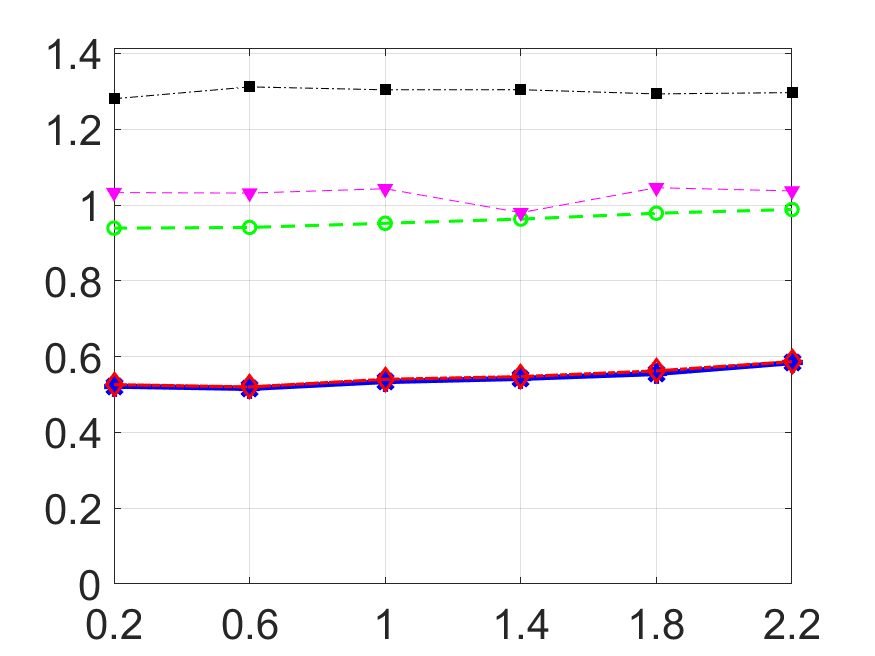}}
    \subfigure[$\theta_{\max}=\pi/8$]{
        \includegraphics[width =0.3\textwidth]{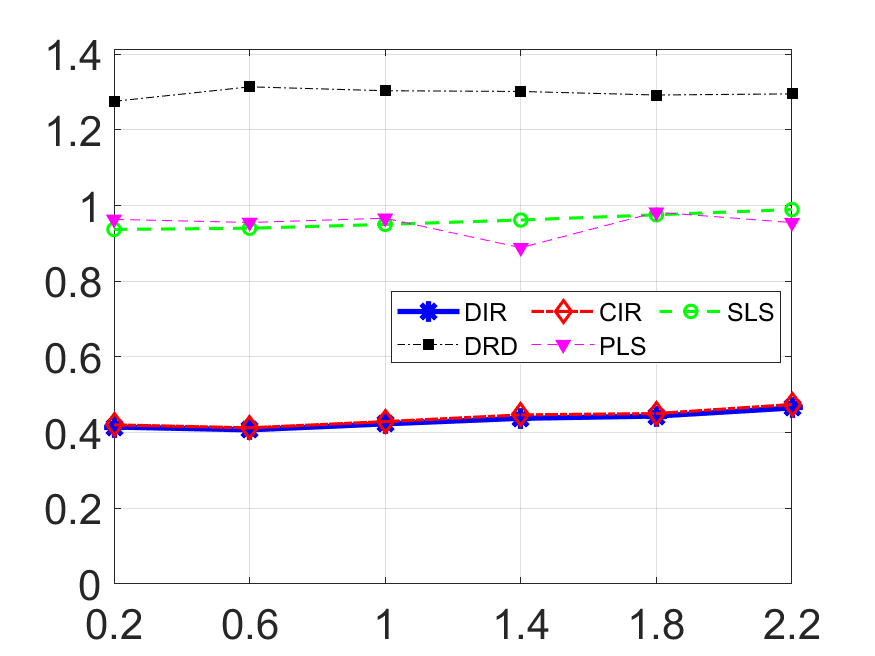}}
    \caption{Average $l_2$-error against the intensity parameter $k$ of noise for the proposed approach DIR (solid-star), SLS (dash-circle), PLS (dash-triangle), DRD (dot-dash-square), and CIR (dot-dash-diamond) under 
    different levels of angular similarity: $\theta_{\max}=\pi/3$ (a), $\theta_{\max}=\pi/4$ (b), and $\theta_{\max}=\pi/8$ (c). The simulations fix $N=2400$, $p=20$ and $m=30$, and let $(\sigma_j,q_j)$ take the values $\{(0.1,0.75),(0.2+0.4\times(k-1),0.125)\}$ with equal probability $1/2$, where $k=1,\ldots,6$ controls the intensity of noise.}
    \label{Figure: Heterogeneity of Noise}
\end{figure}

\begin{table}[!htbp]
    \centering
    \caption{The effect of heterogeneity for noise.}
    \begin{tabular}{c ccccc }
    \hline
        $\sigma$ & method & 1/4-quantile & 1/2-quantile & 3/4-quantile & Improved Ratio\\
    \hline
    \multirow{2}{*}{0.2}    
    & SLS    &0.4109(0.0029)  &0.5838(0.0021)  &0.7316(0.0014)  & - \\
    & Invex &0.7556(0.0020)   &0.8435(0.0006)  &0.8941(0.0004)  &  0.8813(0.0040) \\
        \hline
    \multirow{2}{*}{0.6}    
    & SLS    &0.3986(0.0032)  &0.5783(0.0017)  &0.7277(0.0013)  & - \\
    & Invex &0.7555(0.0018)   &0.8416(0.0006)  &0.8951(0.0004)  &  0.8930(0.0023) \\
        \hline
    \multirow{2}{*}{1}    
    & SLS    &0.3951(0.0025)  &0.5618(0.0015)  &0.7142(0.0015)  & - \\
    & Invex &0.7521(0.0020)   &0.8351(0.0008)  &0.8892(0.0005)  &  0.8977(0.0026) \\
        \hline
    \multirow{2}{*}{1.4}    
    & SLS    &0.3915(0.0019)  &0.5479(0.0021)  &0.6997(0.0014)  & - \\
    & Invex &0.7407(0.0018)   &0.8256(0.0008)  &0.8798(0.0006)  &  0.8903(0.0026) \\
        \hline
    \multirow{2}{*}{1.8}    
    & SLS    &0.3759(0.0024)  &0.5316(0.0016)  &0.6869(0.0013)  & - \\
    & Invex &0.7365(0.0019)   &0.8195(0.0011)  &0.8738(0.0007)  &  0.9000(0.0024) \\
        \hline
    \multirow{2}{*}{2.2}    
    & SLS    &0.3639(0.0021)  &0.5243(0.0020)  &0.6720(0.0016)  & - \\
    & Invex &0.7045(0.0032)   &0.8013(0.0016)  &0.8593(0.0011)  &  0.8860(0.0028) \\
        \hline
    \end{tabular}
    \label{Table:HeterNoise}
\end{table}

\begin{figure}[!htbp]
    % \flushleft %左对齐
    \centering
    \subfigure[$\theta_{\max}= \pi/3$]{
        \includegraphics[width =0.3\textwidth]{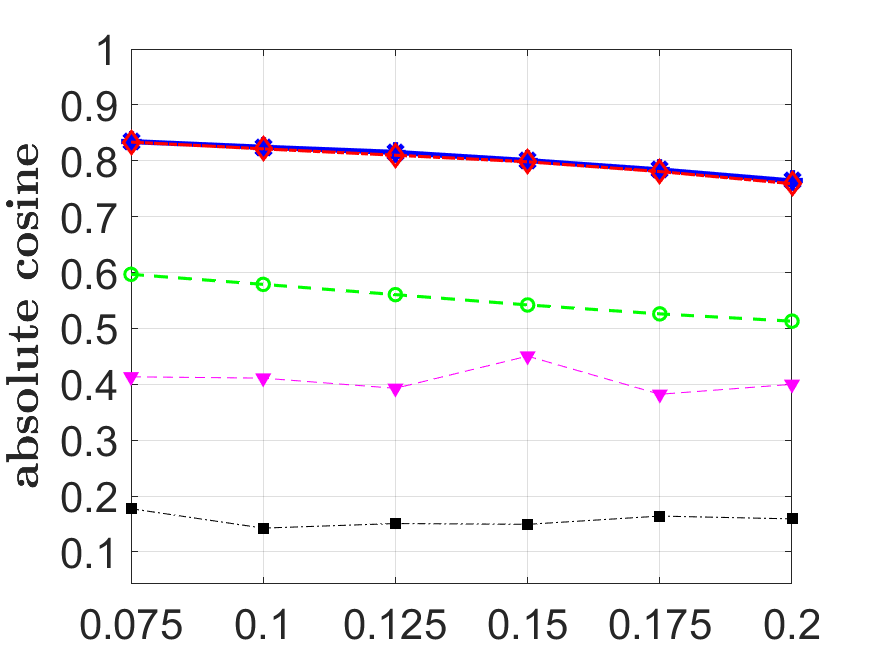}}
        \subfigure[$\theta_{\max}= \pi/4$]{
        \includegraphics[width =0.3\textwidth]{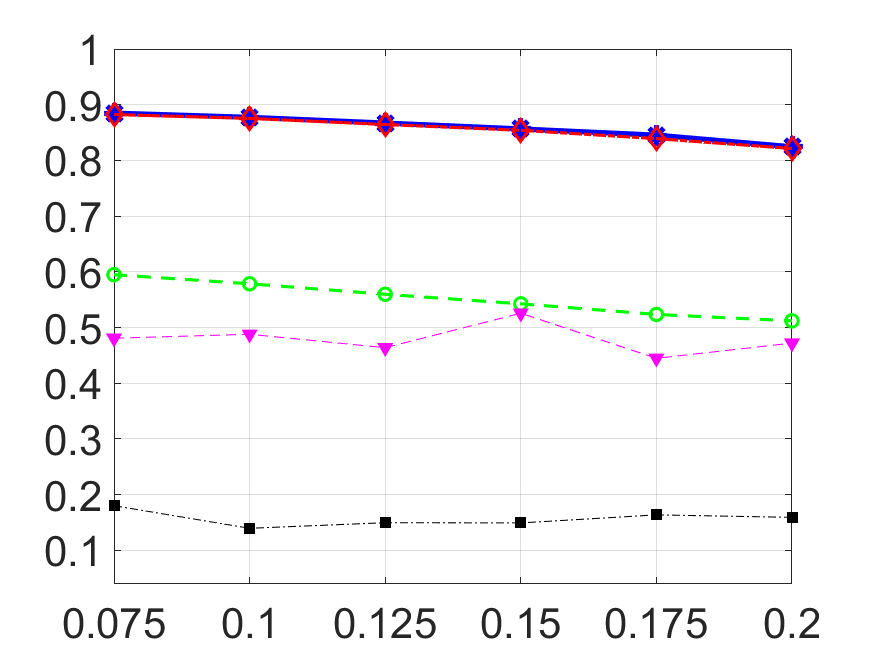}}
    \subfigure[$\theta_{\max}= \pi/8$]{
        \includegraphics[width =0.3\textwidth]{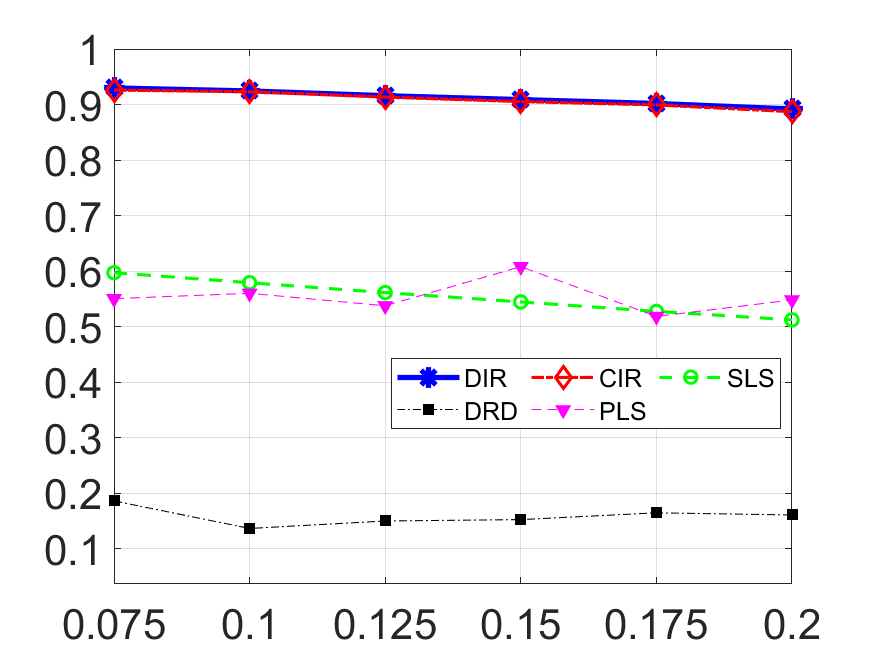}}
    \caption{Average absolute cosine against the degree of sign-flipping probability variation $k$ for the proposed approach (solid-star), SLS (dash-circle), PLS (dash-triangle), DRD (dot-dash-square), and CIR (dot-dash-diamond) under 
    different levels of angular similarity: $\theta_{\max}=\pi/3$ (a), $\theta_{\max}=\pi/4$ (b), and $\theta_{\max}=\pi/8$ (c). The simulations fix $N=2400$, $p=20$ and $m=30$, and let $(\sigma_j,q_j)$ take the values $\{(0.1,0.75),(0.2,0.075+0.025\times(k-1))\}$ with equal probability $1/2$, where $k$ controls the degree of sign-flipping probability variation.}
    \label{Figure:HeterSignCos}
\end{figure}

\noindent{\bf The specific result for sign-flip probability.} In Table \ref{Table:HeterSign}, we present the $1/4$, $1/2$ and $3/4$ quantile of the absolute cosine of nodes for SLS and our method (Invex), with $\theta_{\max}= \pi/3$ and sign-flip probability varying. Improved Ratio represents the ratio of nodes that gain improvement with our method compared to SLS.
Figure \ref{Figure:HeterSignCos} shows the absolute cosine values in these scenarios.

\begin{figure}[!htbp]
    % \flushleft %左对齐
    \centering
    \subfigure[$\theta_{\max}= \pi/3$]{
        \includegraphics[width =0.3\textwidth]{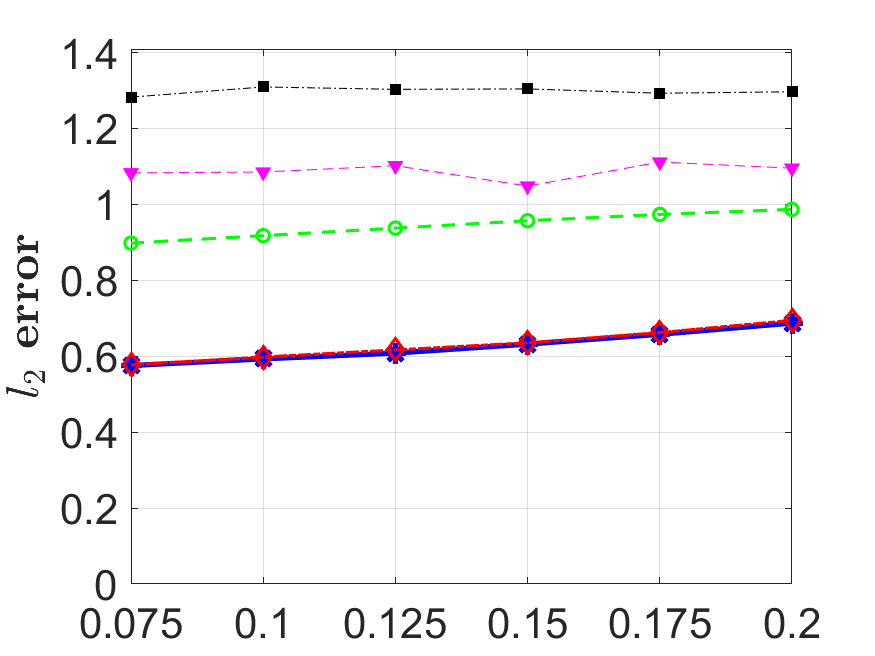}}
        \subfigure[$\theta_{\max}= \pi/4$]{
        \includegraphics[width =0.3\textwidth]{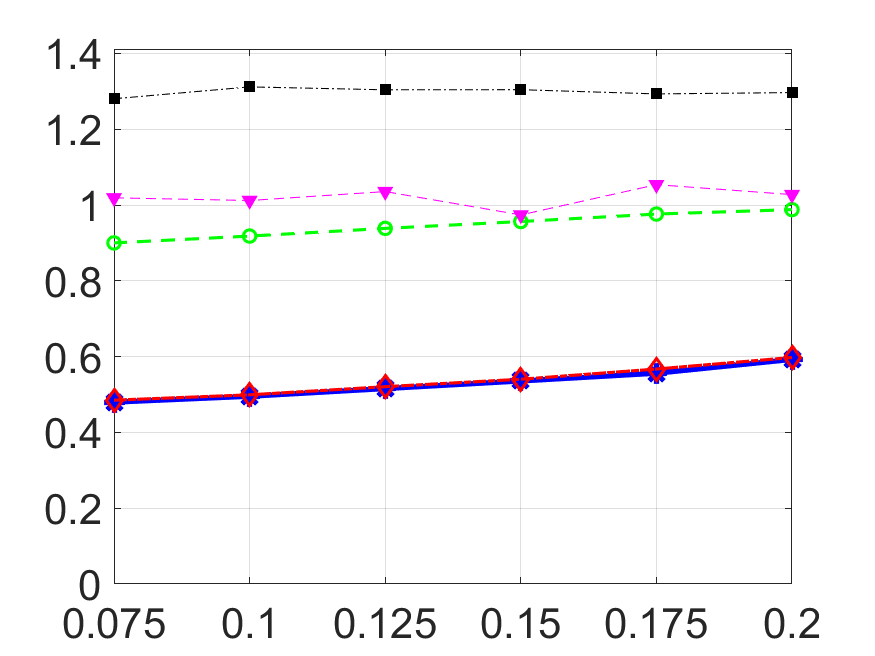}}
    \subfigure[$\theta_{\max}= \pi/8$]{
        \includegraphics[width =0.3\textwidth]{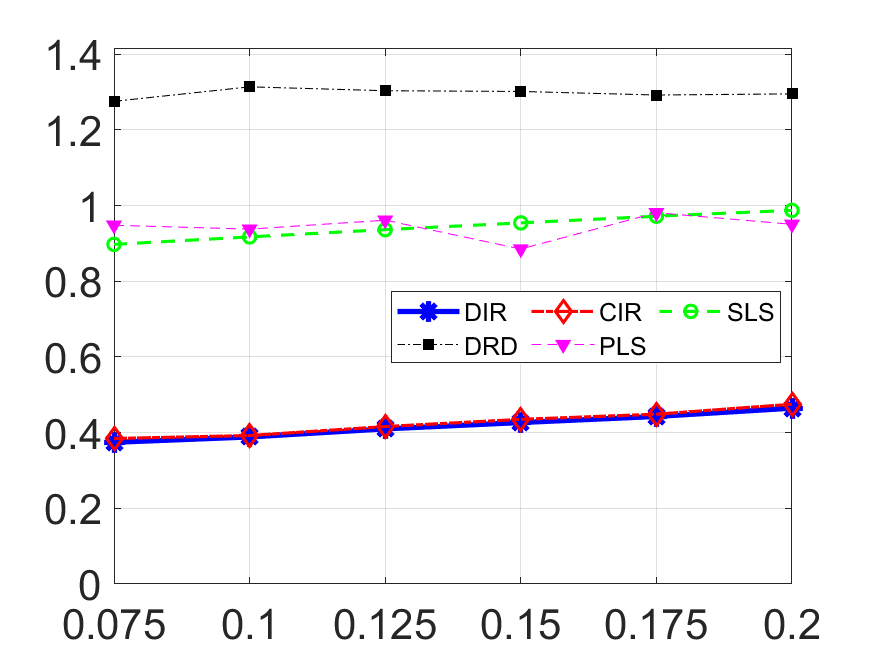}}
    \caption{Average $l_2$-error against the degree of sign-flipping probability variation $k$ for the proposed approach DIR (solid-star), SLS (dash-circle), PLS (dash-triangle), DRD (dot-dash-square), and CIR (dot-dash-diamond) under 
    different levels of angular similarity: $\theta_{\max}=\pi/3$ (a), $\theta_{\max}=\pi/4$ (b), and $\theta_{\max}=\pi/8$ (c). The simulations fix $N=2400$, $p=20$ and $m=30$, and let $(\sigma_j,q_j)$ take the values $\{(0.1,0.75),(0.2,0.075+0.025\times(k-1))\}$ with equal probability, where $k$ controls the degree of sign-flipping probability variation.}
    \label{Figure: Heterogeneity of sign-flips}
\end{figure}

\begin{table}[!htbp]
    \centering
    \caption{The effect of heterogeneity for sign-flip probability.}
    \begin{tabular}{c ccccc }
    \hline
        $q$ & method & 1/4-quantile & 1/2-quantile & 3/4-quantile & Improved Ratio\\
    \hline
    \multirow{2}{*}{0.075}    
    & LS    &0.4355(0.0028)  &0.6234(0.0020)  &0.7775(0.0015)  & - \\
    & Invex &0.7851(0.0018)   &0.8651(0.0005)  &0.9136(0.0002)  &  0.8680(0.0028) \\
        \hline
    \multirow{2}{*}{0.100}    
    & LS    &0.4163(0.0035)  &0.6047(0.0017)  &0.7602(0.0012)  & - \\
    & Invex &0.7712(0.0018)   &0.8560(0.0006)  &0.9055(0.0003)  &  0.8817(0.0023) \\
        \hline
    \multirow{2}{*}{0.125}    
    & LS    &0.4048(0.0027)  &0.5783(0.0015)  &0.7357(0.0011)  & - \\
    & Invex &0.7624(0.0019)   &0.8461(0.0006)  &0.8963(0.0005)  &  0.8940(0.0025) \\
        \hline
    \multirow{2}{*}{0.150}    
    & LS    &0.3952(0.0025)  &0.5527(0.0018)  &0.7088(0.0015)  & - \\
    & Invex &0.7461(0.0019)   &0.8301(0.0009)  &0.8838(0.0008)  &  0.8940(0.0022) \\
        \hline
    \multirow{2}{*}{0.175}    
    & LS    &0.3740(0.0028)  &0.5388(0.0016)  &0.6901(0.0016)  & - \\
    & Invex &0.7302(0.0022)   &0.8174(0.0010)  &0.8739(0.0008)  &  0.8837(0.0029) \\
        \hline
    \multirow{2}{*}{0.200}    
    & LS    &0.3559(0.0032)  &0.5280(0.0018)  &0.6707(0.0015)  & - \\
    & Invex &0.7002(0.0038)   &0.8014(0.0020)  &0.8617(0.0014)  &  0.8837(0.0040) \\
        \hline
    \end{tabular}
    \label{Table:HeterSign}
\end{table}

\subsubsection{Application with dataset SEED}
\begin{figure}[!htbp]
    % \flushleft %左对齐
    \centering
        \includegraphics[width =0.7\textwidth]{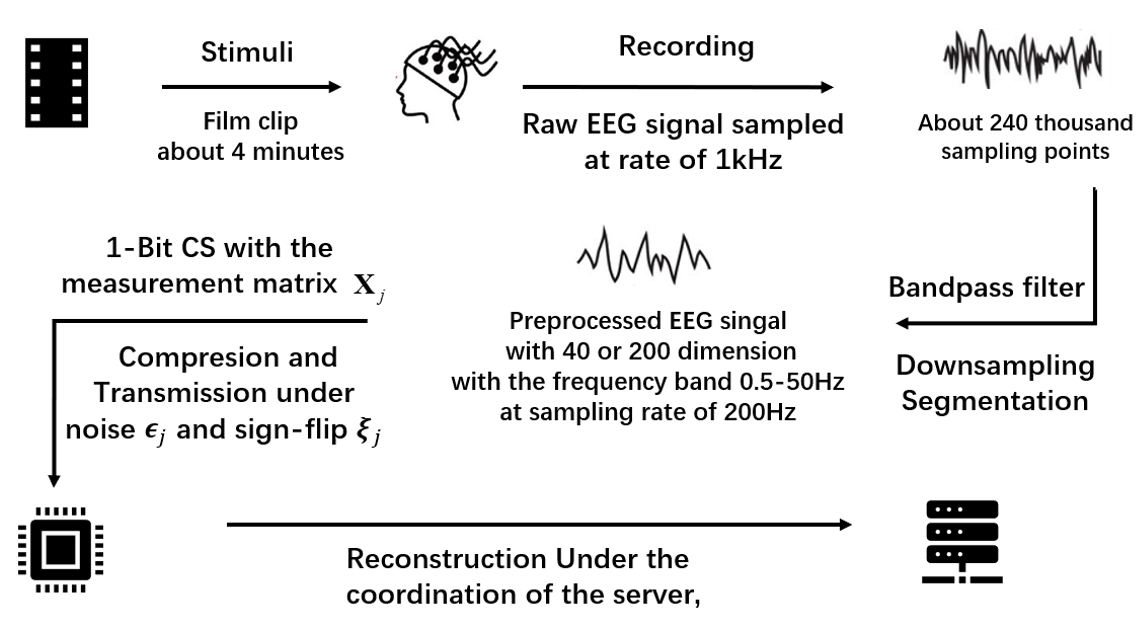}
    \caption{The application process. EEG data recorded during stimulus presentation are preprocessed with downsampling and bandpass filtering. After 1-bit compressive sensing via a measurement matrix, the 1-bit measurements are transmitted to corresponding nodes. Under server coordination, the nodes collaboratively perform signal reconstruction.
    }
    \label{Figure:EmpiricalStudyProcess}
\end{figure}

\noindent{\bf The flow chart for data preprocessing.}
Figure \ref{Figure:EmpiricalStudyProcess} shows the process of our application.

\begin{figure}[!htbp]
    % \flushleft %左对齐
    \centering
    \subfigure[Channel FP1]{
        \includegraphics[width =0.3\textwidth]{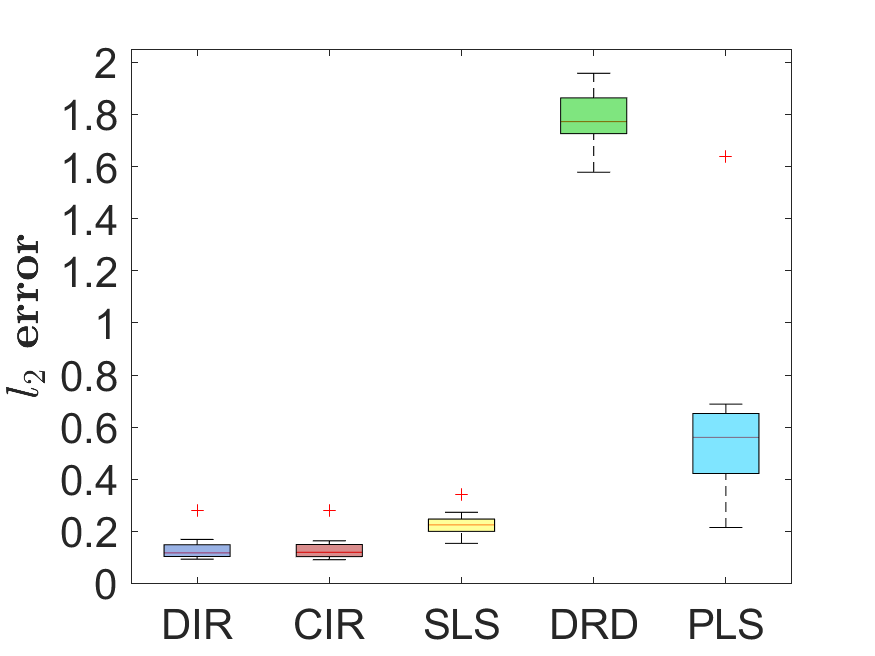}}
    \subfigure[Channel AF3]{
        \includegraphics[width =0.3\textwidth]{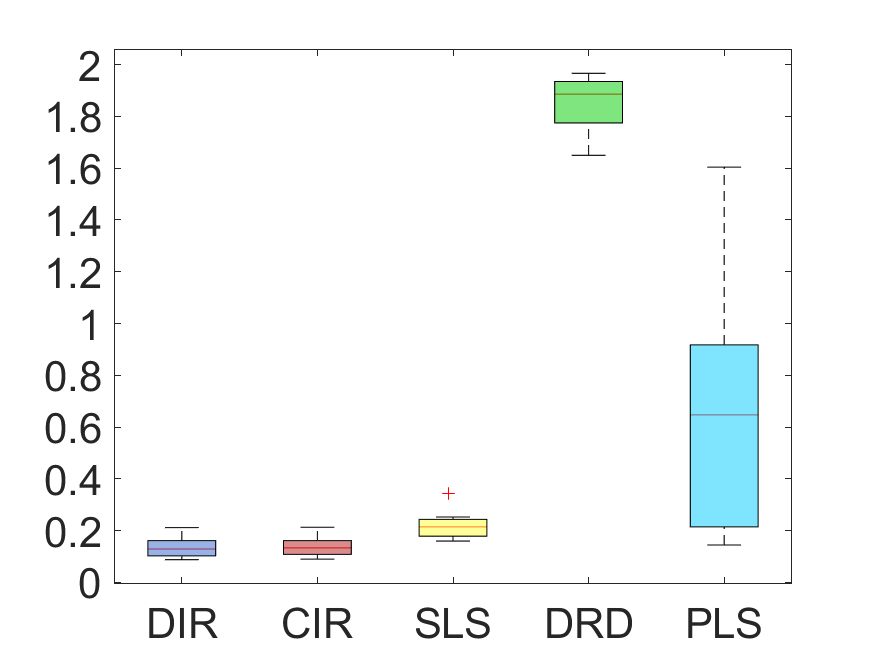}}
    \subfigure[Channel FZ]{
        \includegraphics[width =0.3\textwidth]{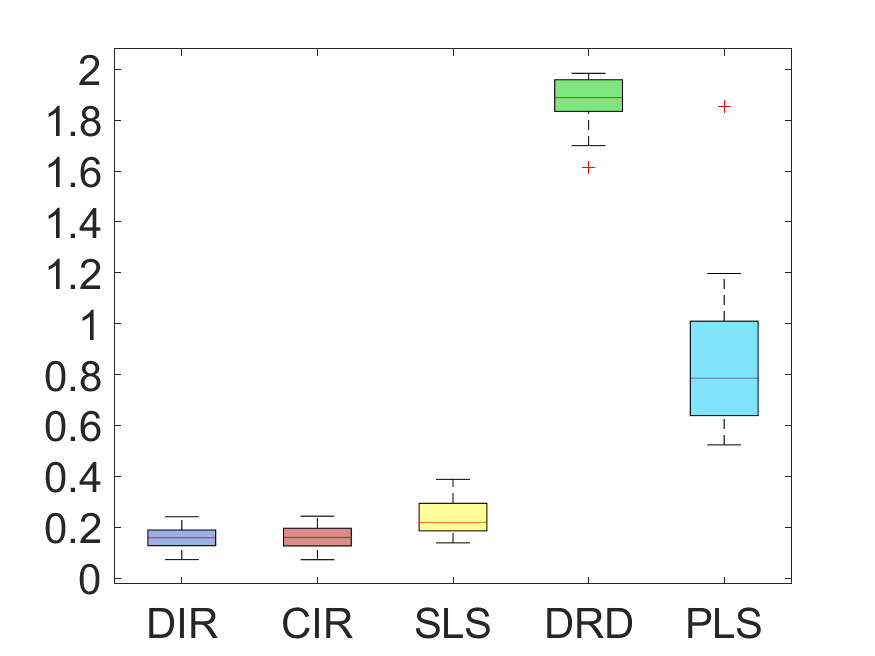}}
    \caption{Distribution of $l_2$-error across nodes for the proposed approach (purple), CIR  (red), SLS (yellow), DRD (green), and  PLS(blue) 
    under different channels FP1 (a), AF3 (b) and FZ (c), with $m=15$, $p=40$ and $n_j=300$ for $j=1,...,m$. The real signal is a $0.2s$ segmented EEG data recorded by above each channel from $15$ subjects under same film clips.
    }
    \label{Figure:EStwoL2SameSample}
\end{figure}

\begin{figure}[!htbp]
    % \flushleft %左对齐
    \centering
    \subfigure[Negative]{
        \includegraphics[width =0.3\textwidth]{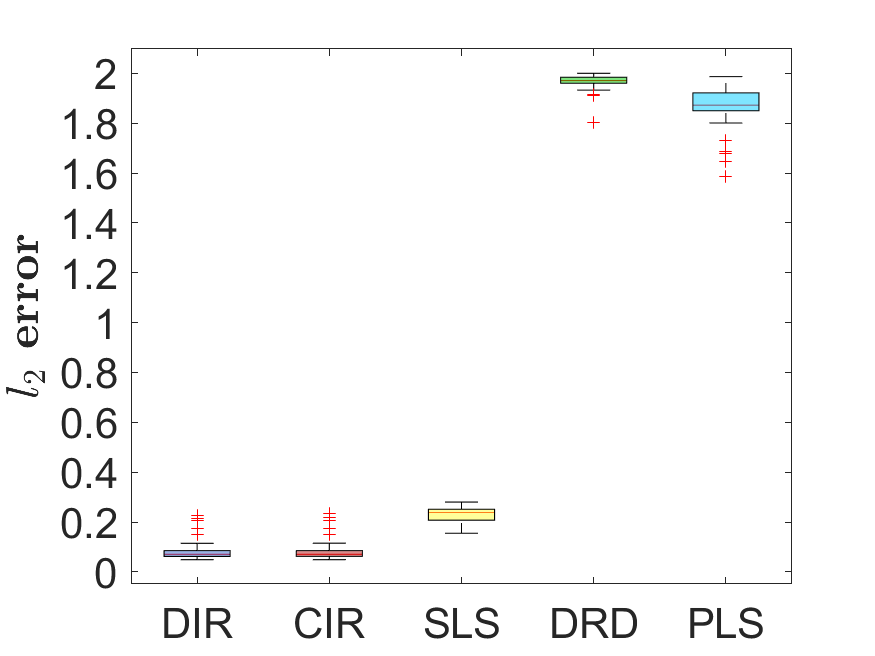}}
    \subfigure[Neutral]{
        \includegraphics[width =0.3\textwidth]{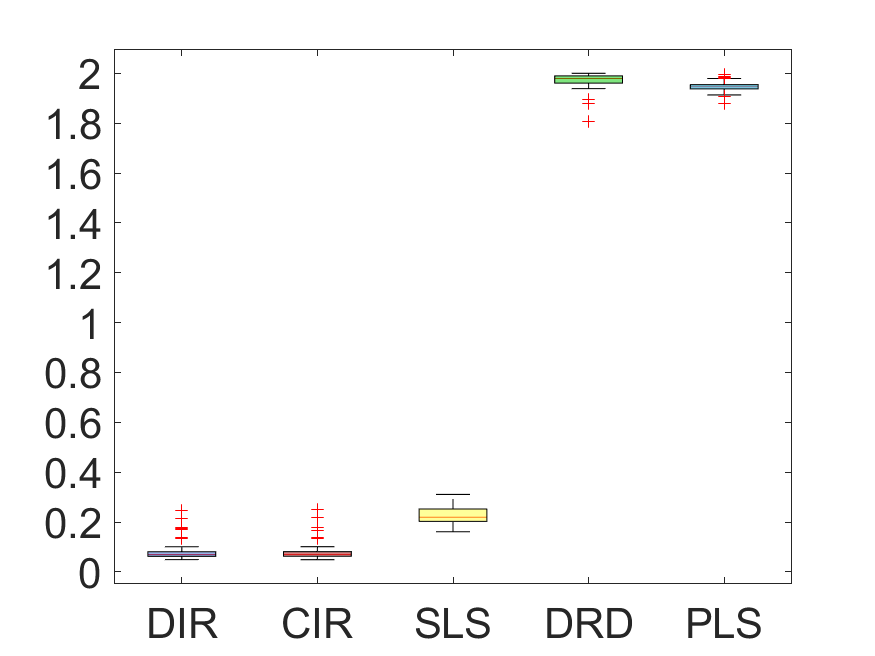}}
    \subfigure[Positive]{
        \includegraphics[width =0.3\textwidth]{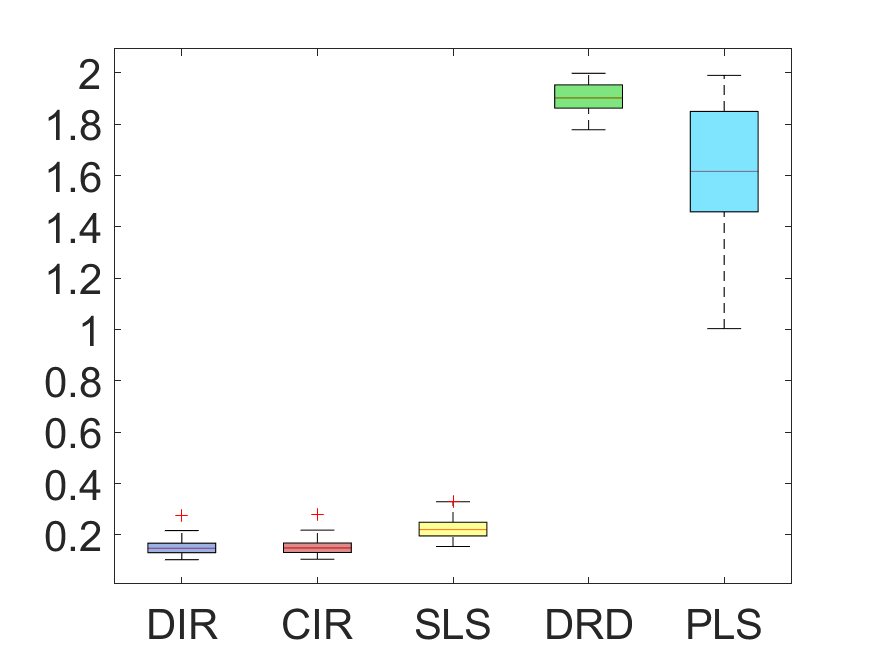}}
    \caption{Distribution of $l_2$-error across nodes for the proposed approach (purple), CIR  (red), SLS (yellow), DRD (green), and  PLS(blue) 
    under different film clips negative (a), neutral (b), and positive (c), with $m=62$, $p=200$ and $n_j=1500$ for $j=1,...,m$. The real signal is a $1s$ segmented EEG data from the same subject under each film clips, recorded by the 62 channel EEG electrode system.
    }
    \label{Figure:ESoneL2SameSample}
\end{figure}

\noindent{\bf The results for the same measurement quantity case.}
Figure \ref{Figure:EStwoL2SameSample} and \ref{Figure:ESoneL2SameSample} show the results for the two experiments in \ref{Figure:EmpiricalStudySketch} with the same measurement quantity across nodes, respectively.

\end{appendix}

%------------------------------------------------------------
% \newpage
{
\centering
\normalem
\bibliographystyle{chicago}
\setlength{\bibsep}{0.5ex} 
\begin{spacing}{1.47}
\bibliography{ref}
\end{spacing}
}